Original, initial version before IntechOpen formatting (with figure number conversion, **see the end**).

# Life is not on the edge of chaos but in a half-chaos of not fully random systems. Definition and simulations of the half-chaos in complex networks

Andrzej Gecow[1]
*andrzejgecow@gmail.com*

**Abstract**
The research concerns the dynamics of complex autonomous Kauffman networks. The article defines and shows using simulation experiments half-chaotic networks, which exhibit features much more similar to typically modeled systems like a living, technological or social than fully random Kauffman networks. This represents a large change in the widely held view taken of the dynamics of complex systems. Current theory predicts that random autonomous systems can be either ordered or chaotic with fast phase transition between them. The theory uses shift of finite, discrete networks to infinite and continuous space. This move loses important features like e.g. attractor length, making description too simplified. Modeled adapted systems are not fully random, they are usually stable, but the estimated parameters are usually "chaotic", they place the fully random networks in the chaotic regime, far from the narrow phase transition. I show that among the not fully random systems with "chaotic parameters", a large third state called half-chaos exists. Half-chaotic system simultaneously exhibits small (ordered) and large (chaotic) reactions for small disturbances in similar share. The discovery of half-chaos frees modeling of adapted systems from sharp restrictions; it allows to use "chaotic parameters" and get a nearly stable system more similar to modeled one. It gives a base for identity criterion of an evolving object, simplifies the definition of basic Darwinian mechanism and changes "life on the edge of chaos" to "life evolves in the half-chaos of not fully random systems".

# 1 Introduction

This is empirical work[2] using simulation. It concerns dynamics in complex autonomous Kauffman networks that are finite and discrete, shows that current theory used for them, based on Lyapunov coefficient in infinite, continuous space, implies false expectations.

Kauffman (**1993**, **1990**) has considered as a general model of a system an autonomous, dynamic, deterministic, complex, random Boolean networks (known as RBN). The discovery of chaos, order, and phase transition between them in such the networks, allowed to look into this very complex world. Lot of works are based on this model (Nghe *et al.*, **2015,** Serra, Villani, Semeria, **2004,** Kauffman, **1990**, **1996,** Aldana, Coppersmith, Kadanoff, **2003a,** Aldana, **2003b,** Luque, Ballesteros, **2004,** Sole, Luque, Kauffman, **2000,** Kauffman *et al.,* **2004,** Shmulevich, Kauffman, Aldana, **2005,** Serra, Villani, Semeria, **2004,** Iguchi, Kinoshita, Yamada, **2007,** Villani *et al*, **2018**). Now, slowly become aware of new important aspects that we have not yet considered adequately. In this paper, such the way is developed, but considered networks are not fully random and use more signal variants than only two. A new obtained vision is clearly different and more adequate for description of adapted objects than now widely accepted. The discovery of half-chaos is the main new element here, which above all frees from strong limitations in systems modeling imposed by the contemporary vision. The statistical properties of the systems are easiest to investigate for fully random systems and from these, we should have started (like Kauffman did), but the systems we model are usually suited to some tasks and are certainly not completely random.

---

[1] No affiliation for this theme, andrzejgecow@gmail.com ; gecow@op.pl
My personal page is: https://sites.google.com/site/andrzejgecow/home
This research did not receive any specific grant from funding agencies in the public, commercial, or not-for-profit sectors.

[2] The description of the investigation and the arguments for introducing the half-chaos given in this article is necessarily shortened and simplified. A much more extensive description is available in supplement (Gecow, **2016a**, **2017a**) to this article. Earlier, simpler versions of the article are available in preprints (Gecow, **2016b**, **2017b**). The data (programs and its sources, results of simulations) analyzed during the current study are available from the author on any request.
Wider list of abbreviations and new terms is placed on the end.
In this work a few abbreviations that are not standard are used: ***s*** - number of equally probable signal variants.
Network types: ***sf*** - scale-free; ***ss*** - single scale; ***er*** - Erdős-Rényi "random"; ***sh*** and ***si*** are respectively ***sf*** and ***ss*** with 30% removal of nodes.
***tmx*** - maximum number of counting steps of time *t*;
***dmx*** - mean maximal damage *d,* i.e. Derrida equilibrium for chaotic behavior;
***q*** - degree of order, fraction of damage which are a small change of network functioning at tmx, capacity of left peak of $P(d)$.

Lyapunov exponents are the most widely used measures to describe chaotic behavior of dynamical systems, however, to check an adequateness of theory by observation of the behavior of finite dynamical networks it must be defined using its main features expected by the theory. The main characteristic of the chaotic behavior of dynamic systems is a high sensitivity to initial conditions, leading to maximally different effects for very similar initial conditions. A small disturbance is a small change in initial conditions. An effect of such small disturbance is called damage. Distribution of damage size (Kauffman, **1993**) ("size distribution of avalanches" in (Villani *et al*, **2018**)) is then the main feature observed in experiment and expected by the theory that may be compared. It is original; theories using Lyapunov exponents or percolation are derivative. The term 'chaos' is used here in such the meaning, similarly as Kauffman does (see ch.2.3). To fit the current theory the damage distributions should fit a Derrida's annealed approximation model (Derrida, Pomeau**, 1986**) and for chaotic systems - an equilibrium level found in it.

It is commonly believed, that system can be only either chaotic or ordered, but not simultaneously both of them - this is shown to be false. On this believing, a "criticality hypothesis" (critical regions have also been said to be "at the edge of chaos" in parameter space of systems) is formulated (see e.g. Villani *et al*, **2018**). Evolution needs small changes which practically occur only in critical regions in such the systems. Current theory and this believing (see e.g. Villani *et al*, **2018**) are based on the assumption that networks are fully random. However, interesting phenomena concerning life occur in not fully random networks due to natural selection. The current theory of chaos was built for functions in infinite and continuous space, but it is used for finite discrete networks (Aldana, Coppersmith, Kadanoff, **2003a,** Aldana, **2003b**), such a method is an approximation. It loses a few important phenomena present in such the networks, but absent in the infinite and continuous space. Due to such the reasons, expectations of the theory that life is on the edge of chaos can be and are inadequate. Here such phenomena are shown; they need much more complex theory which will not use the assumption of full randomness of network and infinite continuous space, but to build such theory is the next step, which is the task for mathematicians. The description of this experiment in the language of mathematical equations seems to me inadequate and unattainable, and in my opinion useless, but mathematicians may have a different opinion. Programming languages are a natural and appropriate tool for describing such issues. I can share the program, but it is complex. It is not true that the below description of experiment is not exact enough to be repeated by every IT specialist. Therefore it is enough exact to understand by mathematicians too.

Indication of adequate ranges of parameters of a complex "purposeful" (adapted) system describing living, technological or social object is a key for modeling their processes. An important parameter is a connectivity (Turnbull *et al.*, **2018**), which current theory strongly limits. The system can be any, e.g. the solar system is also a system, but usually, in human intuition, the system has to somehow work (therefore above "purposeful"), and despite some changeability, it has to keep its identity. The evolution of the system is a term that reconciles two adversities - variability that is the essence of evolution, and the identity of the evolving object. This is not a philosophical problem, but a particular problem for modeling. In this work a base for solving this problem is found. A good approximation of the system description is a dynamic complex network, although it undoubtedly has many important simplifications. We are just entering this subject and it is difficult for our intuitions to operate on more complex, more adequate descriptions, such as process algebras (Nowostawski, Gecow, **2011**).

Half-chaos is a state of the system that is not fully random, with parameters that make the random system strongly chaotic (hereinafter we will call them "chaotic parameters", such the parameters are usually estimated for real systems), however small disturbances give an ordered reaction (small damage) with a similar probability to a chaotic reaction (damage near the Derrida balance (Derrida, Pomeau, **1986**), Fig.1cd). Acceptance of changes that trigger ordered reactions preserves the half-chaotic state allowing for a long evolution of the slowly changing system (the system retains identity), but acceptance of one change that gives a chaotic reaction leads to practically irreversible entry into normal chaos (the system works completely different, ceases to be itself). Thus, the basic Darwinian mechanism emerges - this has large interpretational consequences.

The assessment of whether a given system is chaotic or ordered is currently based on parameters that in the case of a half-chaotic system indicate chaos for the fully random system (I call them "chaotic parameters"), but the behavior of the system turns out to be inconsistent with such prediction. This work presents half-chaotic systems and simple ways to obtain such systems. The experimental results are unambiguous and easy to repeat. The constraints forming the half-chaotic system are small, which means that there are a lot of such systems, though undoubtedly significantly less than of fully random.

The practical result of this work is the realignment of the acceptable range of parameters for system modeling. This is a fundamental change. First and foremost, "chaotic parameter" for the Kauffman (Boolean) network - connectivity is included in this scope, but also a larger number than two of signal variants, also omitted due to the effect in the form of a chaotic system (for fully random systems). The need to introduce a larger number of signal variants for statistical investigations was already explained in (Gecow, **2011**). The maintenance of the name of the 'Kauffman network' for such a network was there postulated, to be no longer synonymous with Boolean networks. However, these postulates acquire practical significance only after demonstrating half-chaos.

# 2 Main assumptions

## 2.1 Variables *K, k, t, N, A* and *d* in Kauffman networks

The considerations concern the statistical stability of the deterministic discrete Kauffman networks (Kauffman, **1969**, **1990**, **1993**) (a little bit extended). The network consists of ***N*** nodes. A node in such a network receives signals at the ***K*** inputs, converts them uniquely using its function to the output signal called the state of the node, and then sends it to other nodes by ***k*** output links. States of all *N* nodes together creates a state of the network. The calculation of function takes a time step. Up to now, 2 (logical) signal states (variants) have been used. In the simplest case, it was assumed the same probability of signal variants and full randomness of connections, functions, and initial states of each node, such networks were called **RBN** (Random Boolean Networks). Here, deviation from this full randomness is made[3] by assuming **short attractor** (a small number of time-steps until meeting the same network state), especially – **point attractor** (next network state is the same). In other here described investigations (met7, ch.3.5) – by controlled construction of in-ice-modular network (ch.3.3) or (met1-4b, ch.4) – by an increase of the fraction of negative feedbacks or classic modularity. *K* (called "connectivity", see (Turnbull *et al.*, **2018**)) was the basic variable for Kauffman.

Synchronous computing is used, i.e., the states of nodes from the discrete time ***t*** are input signals and arguments of the function of other nodes, and the results of these functions are nodes states at the next moment ($t$+1). Variable $t$ – is the number of time steps from a disturbance initiation. As the disturbance a permanent change in the value of the function of the node for its input state is used at the time $t = 0$; in method '8' (**met8**) it was an addition or removing a node. Parameter ***tmx*** - the maximum number of calculated time steps is chosen arbitrarily, but it is checked whether its increase does not change the results (Figs.9, 6, 7).

Considerations have been limited to autonomous systems – they do not take signals from the environment. Determining the states and functions of all nodes and the connections between nodes uniquely determines the trajectory - consecutive states of the whole network (sets of states of all nodes). We simulated the process of transformation of the disturbed system on the section *tmx*, then we compared the resulting state of the system with the undisturbed system. It is also looked after the node functions are correctly random, but this assumption cannot always be fully met, so the impact of the derogations is checked.

The size of a change in a network function at time $t$ after a small disturbance is measured by the number ***A*** (from Avalanche (Serra *et al.*, **2007**)) of the nodes, which have a different state in the pattern network – identical network, but without disturbance. The value $d = A/N$ is called damage. The distribution of damage size at the time *tmx* as ***P(d)*** or ***P(A)*** is an especially important result (Fig. 2).

For random networks, this result creates two system states – ordered and chaotic. In system parameters space they occupy areas which Kauffman calls 'solid' and 'gas' respectively. Between them, there is a fairly quick transition (near $K = 2$, if Boolean signals are equally probable) treated as a phase transition. Only in systems in the vicinity of this transition (Kauffman calls it 'liquid' - the area between 'solid' and 'gas') changes in the system function (damage) often enough are small, therefore suitable for biological evolution. This is the main basis for the Kauffman's hypothesis: life on the edge of chaos. However, this conclusion aroused doubts (Gecow, **2011**), therefore, it has been subjected to a deeper analysis presented here.

The conflict (Aldana, S. Coppersmith, L. P. Kadanoff, **2003a**, Aldana, **2003b**) of a size of *K* in the Kauffman model and *K* estimated from nature (Turnbull *et al.*, **2018**) is a problem solved here. Kauffman postulates that the natural property of the random ordered systems (order for free (Kauffman, **1996**)) is the source of stability, but then *K* should be extremely small ($K \leq 2$) (Derrida, Pomeau, **1986**). The attempts to prove that the real genetic network (using model **GRN** – Gene Regulatory Network) is ordered (Serra *et al.*, **2007**, Shmulevich, Kauffman, Aldana, **2005**, Serra, Villani, Semeria, **2004**, Villani *et al*, **2018**) assume such a source of stability. Different circumstances allowing system with greater *K* to be in the ordered phase were indicated (p.48 in (Aldana, S. Coppersmith, L. P. Kadanoff, **2003a**)), such as a significant difference in probabilities of logical states (Derrida, Pomeau, **1986**), or deviation from the randomness of the function (canalizing (Kauffman *et al.*, **2004**)), but these and other suggestions are not satisfactory for many reasons (Gecow, **2011**). The model GRN has disappointed many expectations, mainly due to restrictions arising from the range of 'liquid region', it was replaced by the more attractive Banzchaf model (Banzhaf, 2003), but GRN is still being studied (Kinoshita, 2019).

---

[3] It is made using few method, in short: 'met'. Each of them is called using digit on the end and, if need be, some letter for its variant. In this case they are: met4c, met4d, met5, met6, met8, described in ch.3.1 - 4.

For investigation shown here, as typically, the same $K$ for all $N$ nodes of the network are taken.

## 2.2 More than two signal variants $s \geq 2$

According to my previous (Gecow, 2011) suggestions, here I also study a larger number of ***s*** (> 2, usually 4) of equally probable signal states, which in random networks for every sensible $K$ ($\geq 2$) always gives chaos (Fig.1). In the range of sensible parameters $s$ and $K$, the order appears only for $s = 2$ and $K = 2$, it is absolutely exceptional (Fig.1bd). Attempts to introduce more signal states already exist (Luque, Ballesteros, 2004, Sole, Luque, Kauffman, 2000), but they assume the possibility of an ordered phase for the random network therefore these states cannot be equally probable.

I repeat here briefly my basic arguments given in (Gecow, 2011 ch.2) for using $s \geq 2$ in Kauffman networks for statistical investigations:

1 - Using Boolean network we can describe each complex relationship (mechanism), but bringing to two-value description frequent cases where significant signals take more than two variants, we generate unrealistic situations, presumably - to skip. In the statistical analysis, however, they are not skipped and give a false picture. Or we simplify something which we do not want to simplify. In both cases the statistical investigation is false. It was shown on the example of the thermostat (Gecow, 2011 Fig.3, p.292). The only way is to use a real number of signal variants and not limit ourselves to only two Boolean alternatives. Because such case is frequent, then in one system it should appear for a lot of signals and the investigations based on $s = 2$ must be false (extreme).

2 - Two variants are often subjective (Gecow, 2011 ch.2.1.2). There are typically lots of real alternatives, but we are watching one of them and all the remaining we collect into the second one. Typically our interesting variant has much lower probability, resulting from the similar probability of each one, but description basing on ***p*** – the probability of signal variant leads to much different results (Fig.1a) than for the case, when we consider all variants. Adding the parameter $p$ to the two-value description does not solve the problem here. Adoption of $s \geq 2$ equally probable signal variants is an alternative method of model realignment. However, it seems to be often more adequate. Both methods give different results which significantly increases the importance of the correct choice of description.

3 - Parameter $s$ is more important in the description of damage spreading than $K$ treated as the most important – see Fig.1b. Value ***dmx*** – mean maximal damage, i.e. Derrida equilibrium for chaotic behavior (Fig.1cd), much stronger depends on $s$ than on $K$.

## 2.3 Criteria of chaos, coefficient of damage propagation $w$

The main characteristic of the chaotic behavior of dynamic systems is high sensitivity to initial conditions, leading to maximally different effects for very similar initial conditions. It is original, theories using Lyapunov exponents or percolation are derivative. I use the term 'chaos' in such the meaning, similarly as Kauffman (Kauffman, 1993) does. For chaotic Kauffman networks a small initiation of damage typically causes a large avalanche of damage which spreads onto a big part (percolates) of the discrete and finite system and ends at a Derrida equilibrium level *dmx* (Derrida, Pomeau, 1986, Gecow, 2011, Derrida, Weisbuch, 1986) (Fig.1cd), which is a maximal loss of information about the previous system. Such distance cannot be infinite in the finite network witch finite discrete $s$. The existence of this limitation is the main difference between this 'chaos' and the more commonly taken definition (Schuster, 1984) used for continuous variables on infinite space, where Lyapunov description works. **The term 'chaos' is not reserved for one of those separate areas. The distribution of damage size is the experimental base to classify a particular system of Kauffman network as chaotic or ordered using levels of damage equilibrium calculated from Derrida's annealed approximation** (Fig.1d). In Derrida's model only case $s,K = 2,2$ (I use $s,K$ as a vector) is ordered – it has no other cross of diagonal than in $d_{t+1} = d_t = 0$. In any other cases, such cross called here ***dmx*** exists; it is Derrida equilibrium level of chaotic reaction for the disturbance.

In a typical case, the chaos is indicated by Lyapunov exponent, which describe the growth of distance for two, near, initial states. For finite discrete networks, it corresponds to "coefficient of damage propagation" ***w*** described in (Gecow, 2011 ch.2.2.1) and earlier, or eq. 4.8 in (Serra *et al.*, 2007). $w = \langle k \rangle (s\text{-}1)/s$. It can be treated as damage multiplication coefficient on one node if only one input signal is changed. It indicates how many output signals of a node will be changed on average. For an autonomous network with fixed $K$, $\langle k \rangle = K$ and we can use $w = K(s\text{-}1)/s$. It is easy to see that for $w > 1$ damage grows, for $w < 1$ it disappears and $w = 1$ is critical – for $s = 2$ it gives known critical $K_c = 2$. In (Aldana, S. Coppersmith, L. P. Kadanoff, 2003a) similar equation (6.2): $K_c(s\text{-}1)/s = 1$ is given which is a case for the condition $w = 1$. Coefficient $w$ is a simplification for the beginning of damage growth, later a case of more than one changed input signal happens more and more often, but this first period is crucial (Fig.1c).

Note, we are going to know: is a particular network chaotic, ordered, or something else, therefore we test it by statistical experiment. We make small disturbance (perturbation) and look how great is a change of a function (damage *d*) of this deterministic network comparing to undisturbed network. Damage is an effect of this small disturbance. We make a lot of such small disturbances (see Fig.6), each in the same network being tested, and we get distribution *P(d)* for one, tested network. For a chaotic network, the *P(d)* contains one peak near *dmx*, for ordered – one narrow peak near $d = 0$. If there are both the peaks in the distribution for one particular network, then it is neither chaotic nor ordered network, it may be half-chaotic.

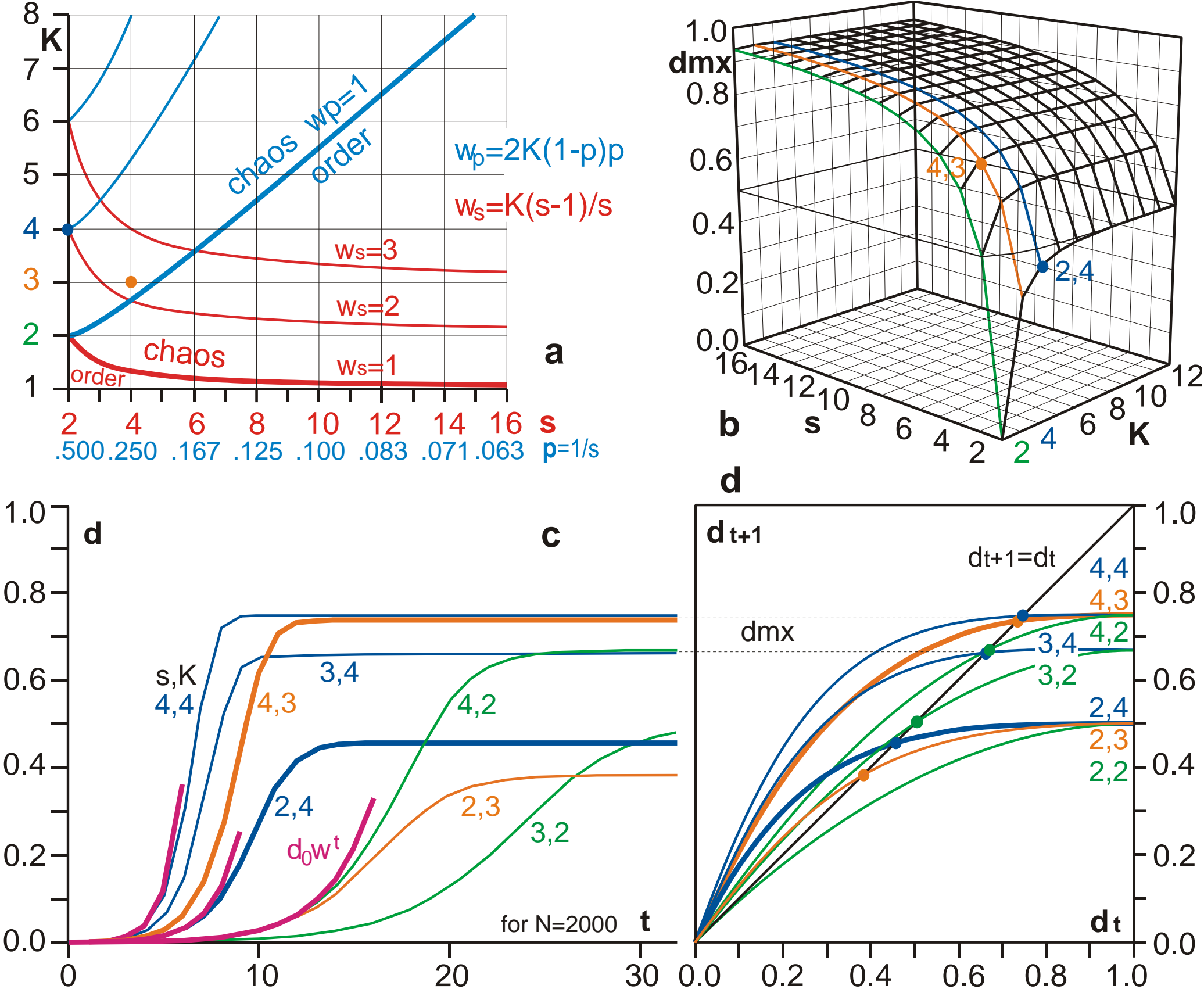

**Fig.1 [1] Comparison of models based on *p* and *s;* - of influence of *s* and *K* on Derrida equilibrium, Derrida plot; *d(t)*.**

**a** - Comparison of models based on probability *p* of one Boolean signal variant and on *s* equally probable signal variants in dependency on *K*. As the basic argument *s* is taken. For it *p* is added as 1/*s*, it is for the case if in reality there are *s* different equally probable signal variants, but we are interested only in one of them, and rest we collect to the second one. Values of the coefficient of damage propagation $w_s$ for *s,K* and $w_p$ for *p* and *K* are used. The equation for $w_p$ is taken from (Kauffman, 1990, Shmulevich, Kauffman, Aldana, 2005). Both models give very different results, it means, that they cannot replace each other. (See also (Gecow, 2011 Fig.4)).

**b** - Derrida equilibrium (*dmx*) for chaotic response in the system of *s,K*. Kauffman using Boolean networks has considered only *K* as the most interesting variable, but *s* influences *dmx* more hardly. However, he cannot use *s* other than 2, because for each s>2 the chaos is present ($dmx > 0$ exists), like for any $K > 2$. Among sensible *s,K*, only for 2,2 exist order, it is an especially extreme case.

**c,d** - Theoretical damage spreading calculated using the Derrida's annealed approximation model.

**d** - The change of damage in one step of the time in synchronous calculation known as the 'Derrida plot', extended (Gecow, 2011) for the case $s > 2$. The crossing of curves $d_{t+1}(d_t, s, K)$ with diagonal $d_{t+1} = d_t$ shows equilibrium levels *dmx* up to which damage can grow. Case $s,K = 2,2$ has a damage equilibrium level in $d = 0$. These levels are reached on the left which shows damage size in time dependency. For $s > 2$ they are significantly higher than for Boolean networks. All cases with the same *K* have the same color to show the influence of *s*.

**c** - In this plot expected *d(t)* for $N = 2000$ is shown. It is an effect of 'Derrida plot' shown in d. A simplified expectation $d(t) = d_0 w_s^t$ based on coefficient $w_s$ is shown for the first critical period when *d* is still small - three short curves to the left of the longer curves reaching equilibrium.

Parameter ***s,K*** (treated as a vector) is the main variables in the simulations. Most of the studies are made for *s,K* = 4,3, also sometimes for *s,K* = 2,4 (that is, for Boolean network). They provide highly chaotic random systems - 'coefficient *w* of damage propagation' is significantly higher than one.

### 2.4 Types of networks

Several **types** of networks are considered. They differ in the rules of their creation (for *sf* and *ss* see Fig.2 in (Gecow, **2011**)) and distributions of *k* (output links), (*K* – input links is fixed for all nodes of particular network): ***sf*** (scale-free (Barabási, Albert, Jeong, **1999**)), ***er*** (classic Erdős-Rényi (Erdős, Rényi, **1960**) "random"), and ***ss*** (single-scale). In the figures, the second letter of these shortcuts indicates the network type. In studies **met8** (denoted in figures by '8') the network grew - an addition or removal of the node was the disturbution. There networks ***sh*** and ***si*** are respectively ***sf*** and ***ss*** with 30% removal.

Parameters: network **type** together with ***s,K*** (treated as a vector) are the main variables in the simulations. In a wider description (Gecow, **2016a**, **2017a**) of here presented investigation, I used more network types.

### 2.5 The main results

At the beginning, in ch.2.1 there is the statement: 'the distribution of damage size at the time *tmx* as ***P(d)*** or ***P(A)*** is an especially important result'. It is shown in Fig.2 for the main range of investigation and in Fig.10 for mechanisms supporting half-chaos. However, it is the base for more important conclusions.

In obtained here distribution of damage size for the particular system there are two peaks: the left of small changes (ordered behavior) and the right of big changes (chaotic, near Derrida balance). Sharp boundaries of these peaks, supported by a clear gap between them define a "**small change**".

The main result, however, is a "**degree of order**" ***q*** – a fraction of effects (damage) of small perturbations which fit into the range of the "**small change**" of the functioning at the time *tmx*. It is summarized in Fig.8 and Fig.10d. This *q* corresponds to the contents of the left peak or probability of acceptance of changes in the modeled evolution (lack of elimination).

The **degree of order *q*** is the base (see ch.2.3) to state, that we found half-chaos using definition given in the Introduction: **Half-chaos is a state of a system that is not fully random, with parameters that the random system make strongly chaotic, but small disturbances give the ordered reaction with a similar probability to the chaotic reaction.** Such state is contrary to the current view, but the current view is based on the assumption of full randomness of the network which typically is not fitted.

The "**small change**" is a criterion of the acceptance of perturbing permanent changes creating the evolution, which is enough (Fig.3) to stay in half-chaos. It is the **evolutionary stability of half-chaos**. It **was included in the half-chaos definition**. **Acceptance of one** perturbing permanent change that gives a **big change** of the functioning at the time *tmx* (chaotic reaction) **leads to practically irreversible entry into normal chaos** (elimination)**.** Note that in such great change of behavior only states of network nodes differ before and after, but in both cases they have the same, random–look distribution. Nothing has changed for currently used methods to define: is this network chaotic or ordered, but the behavior is absolutely different.

## 3 Half-chaotic systems, construction and mechanisms

### 3.1 Short attractor and secondary initiation as the main mechanism

The preliminary search (met1-4ab described in ch.4) of mechanisms enlarging stability for chaotic systems allowed for a deeper look at the process and its determinants. However, it turned out that they concern mechanisms of secondary importance which only support the main mechanism based on a short attractor effected from the phenomenon of secondary initiation. The first initiation does not have to lead to quick explosion to chaos, it even could fade out. Secondary initiation - the cases of re-appear at the inputs of disturbed node its initial inputs state for which the function has been permanently changed are responsible for the decline of *q(t)* with increasing *t* (Figs.6b, 9). Such a secondary initiation takes place in different conditions than the previous one and can also lead to entering chaos or fade out. After a round of attractor new such cases are no longer present (see Fig.5ab). If up to this point explosion to the chaos does not take place, then it will not appear later. For short attractor, it can happen with not a negligible probability. To check it, *tmx* must be greater than the sum of length (in time steps) of attractor and path to the attractor. In below-described researches, it turned out that global attractor can be large if it is assembled of few independent short local attractors, which is a typical case for in-ice-modularity (ch.3.3).

## 3.2 System with point attractor is half-chaotic

The study of **the systems with a point attractor** (further – 'point attractor system', system state is not changed over time *t*, attractor is extremely short, length = 1), with parameters *s,K* = 4,3 (**met4c**) and 2,4 (**met4d**), (see more in ch.4.3, ch.3.3 and description of Fig.10) which make random systems highly chaotic, gave clear results - such systems **are neither ordered, nor chaotic**. Both reaction variants on a small initial perturbation (ordered - a small change in the functioning and chaotic - a big change nearby of Derrida equilibrium - Fig.2) appear in similar proportions (Fig.8). **This state was named "half-chaos"**. In this state, the resultant change in the functioning (damage) can be either very small or very large (explosions to the chaos Fig.6), but almost no intermediate changes (Figs.2, 6, 7, 10). This defines a **small change** in a natural way. There remains the problem of the length and condition of the evolution of the half-chaotic system.

**Obtaining a point attractor is simple**, just after the random generation of networks (nodes connections and functions) and the states, it is enough to take that for the current state of the node inputs a node function gives the current node state. For the remain states of the input - functions stay random. The point attractor system in Kauffman terms is a completely frozen system – there is only "ice" (nothing changes). The predominance of the ice is a spontaneous property of ordered systems. Obtaining small change after disturbation of half-chaotic, point attractor system, we can expect "a small lake of activity in the ice," (originally (Kauffman, **1990**): "unfrozen islands"), which is the essence of the 'liquid' area of random systems, where Kauffman sees place for life. But such a system ceases to be a point attractor system. It turns out that the vast majority (typically over 99%) of "small changes of functioning" gives also point attractor systems. Therefore, evolution may be long, however, such the model is quite extreme and unattractive.

Simulation studies and their analysis include many important details that are unfeasible to include in this article. They are described in more than 170 pages of the report (Gecow, **2016a**), Only basic ones will be listed here. The particular system is calculated at *tmx* discrete time steps *t* after disturbation, and then at *t* = *tmx*, more adequate value for the final results (Figs.2, 8) is recorded as averaged *A* over the last 50 counting steps *t*. Due to the strong influence of various factors often sporadic, **formal errors in the obtained results are not calculated, judging such a calculation as clearly inadequate and misleading**. This problem is limited to the similarity of results from the similar simulations and the visual evaluation of fluctuations. Given here a number of networks in described series of simulations concern showed results, but often experiments were repeated in a similar way, giving a much greater certainty.

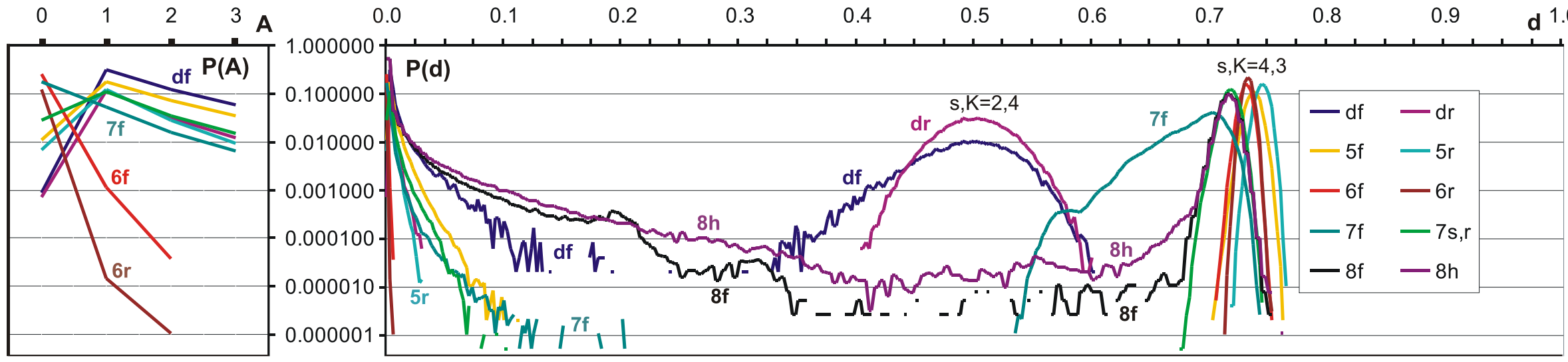

**Fig.2 [7] The main result – distribution of damage size.**
Symbol of the method begins a signature. The methods: 'd', '5', '8' start from point attractor; 'd' (met4d, see ch.4.3) is the only with *s,K* = 2,4 and without evolution, remain *s,K* = 4,3 with evolution; '6' (met6) starts from small attractors; '7' (met7ea) starts from constructed in-ice-modular system. After the method the second letter of network type ends signature. Results presented here (except 'd') are a sum from 4 already stabilized sets of initiation (see Fig.3). The gap between the peaks - left (ordered) and right (chaotic, near Derrida balance, different for *s* = 2 and 4) is not empty only for not really small disturbations by adding or removing a node ('8' – met8). The share of the left peak as *q* – degree of order is summarized in Fig.8. It is the basic result of this study; it allows to introduce half-chaos. Collecting only permanent changes which give damage from the left peak (i.e. small changes) is sufficient to keep half-chaos in the evolution (Fig.3). The shape of the left peak is important for the modeling an evolution of adapted systems. It is shown (without '8') in more details on the left for variable *A* = *d***N* where *N* is = 400. In the experiment '6' there is practically only *A* = 0 due to lack of in-ice-modularity. Network *sf* of '7' differs from the others in the left slope of the right peak, (see also Fig.3c) mechanism of this is unknown.

### 3.3 The evolution from the point attractor

Next, for models b (see ch.4.3) and c of the met4 started from point attractors we checked how long can be evolution if it accumulates small disturbances caused small changes of functioning (small damage), but it does not allow new point attractors (**met5**). We received (also in **met6-8**) that it allows to any length of maintenances of the half-chaotic state and stabilizes its parameters (Fig.3). It is the **evolutionary stability** of half-chaos which **was included in the half-chaos definition**. The system still has a **significant prevalence of ice** (Fig.3c), and there are usually some **"small lakes of activity" forming "in-ice-modules[4]".** Among the methods used to check the presence and properties of the in-ice-modules (see also Figs.7cd, 4), the most effective was to **track periods of node states**. The set of nodes with the same period in the process ended of accumulation was treated as a **local cluster corresponding with in-ice-module**. On average, at the same time occurred about 2 local clusters (Fig.3e). In the evolution, sometimes after many in the meantime accumulated changes, there appeared local clusters very similar in terms of nodes composition - a **collection of such local clusters is treated as a global cluster**. Methods to identify global clusters are very complex due to the wealth of different circumstances, including merger and disintegration of global clusters during evolution. However, we can say that they are generally quite stable formations, though they often disappear (freeze) and reappear, often in the other company of remaining global clusters, often changing period. Their average number for a set of initiations presents Fig.3d.

It should be emphasized that the **structure of the nodes connections in the investigated networks was constant and random**, although the randomness had various formulas that define the type of the network. **In-ice-modules are also the classic modules**, however this is only one, supporting, but less important factor. **The main property of the in-ice-modules is the activity** - changes of the states of nodes forming the in-ice-module. **The ice** (the area where the nodes do not change their states) surrounds them and **isolates from the other in-ice-modules**. In-ice-modules are the result of the functioning defined by the functions and states of the nodes in a given structure. Despite the selection of functions for obtaining initial point attractor state, **functions and states of nodes had truly random characteristics**.

Simulations met4, met5 and met8 start from the system with point attractor. **In the met4** (see also ch.4.3) networks *sf* and *er* were tested. Number of nodes $N$ = 400 and 4000, section $tmx$ = 200 and 2000 (no variant $N$ = 4000, $tmx$ = 2000). One set of initialization was tested - for $s$ = 2 (met4d) each node is able to one initiation, for $s$ = 4 there was 3 of the remaining function values. There were gained 48,000 events for each of the three variants of ($N,tmx$). The differences in the results of these variants were not significant (Figs.2, 10), for further research in met5 we used $N$ = 400, $tmx$ = 1000.

We limited **met5** (and next met6, 7, 8) to $s,K$ = 4,3, but these studies were much more complex. For a long process of evolution (accumulation of initiating permanent changes, which give small damage) we were studied many full sets of initiations, therefore the same change in function as an initiation has been repeated, but it was separated by many accumulations. Full evolution of the particular network is a collection of 20 sets ($M$) of initiations after one initial ($J$ in Figs.3, 5, 6) set. In most of these sets, **retrogressive changes were blocked**. This results in the exclusion of a large number of initiations from the measurements and leads to a significant slowdown of evolution. After several such sets, the reversal is allowed (*M1, M7, M13, M19, M20*), assuming that the change has already another circumstance. It also allows to correctly measure of various phenomena that illustrate evolution (Fig.3). Since the attractor is decreasing spontaneously, making it difficult to move away from the point attractor, it is also forbidden to reduce the global attractor to less than 7, and in the *M20* (in met5, 6, 7) to reduce the attractor.

Parameters $q$ and average time of five the latest "explosions to chaos" are the most important, they demonstrate in Fig.3ab lack of converging into chaos. They stabilize starting from set *M7*, despite a slightly elevated length of global attractor was forced. There were happen that the conditions for the attractor size block further evolution. Such processes were interrupted, however, in the main series (of met5 and met6) 100 networks were obtained, which reached the end of *M20*.

It turned out that the amount of a shift (in the range of 2-50) of the point of process start (place of the initiation) after each accumulation is an important factor. We assumed a shift of 50 steps. The study was much broader and deeper, their wide description can be found in (Gecow, **2016a**). Additional attempts of evolution referral more towards the boundaries of chaos gave no noticeable nearing - a condition of acceptance of a small change is enough for any long evolution - gives evolutionary stability of half-chaos.

---

[4] Term "semi-module" was initially used in (Gecow, **2016a, 2017a, 2016b, 2017b**) (which now reach together over half thousand downloads), later (due to finding its connection to classic module) it was changed to "ro-module", next to "ice-module". Now, in the first normal publication, I prefer final name "in-ice-module", which better describe its mechanism strongly connected to the ice as a lake of activity in the ice.

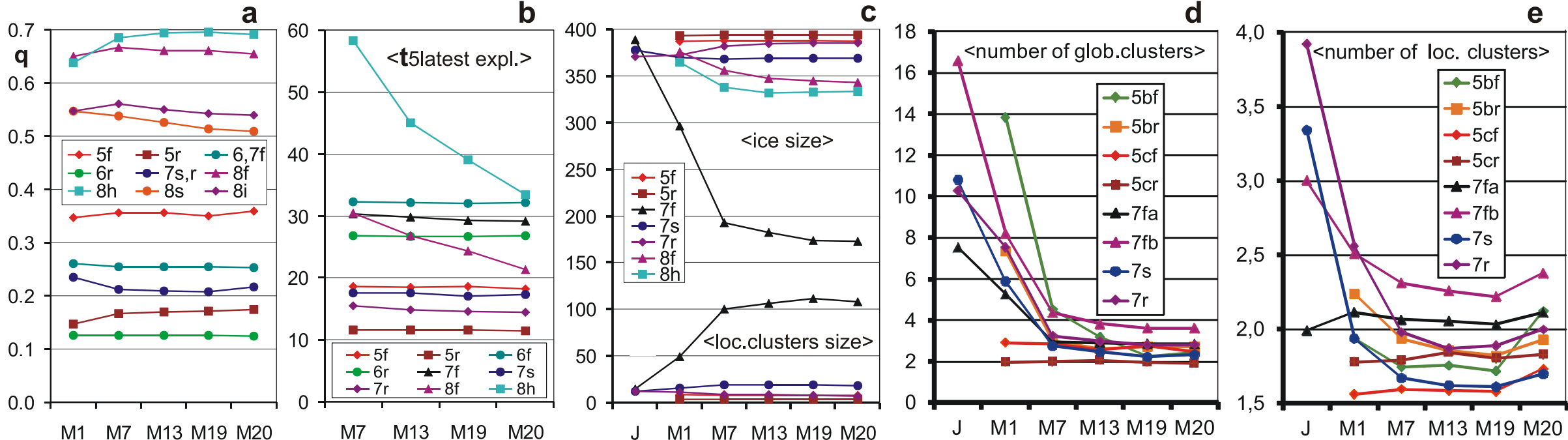

**Fig.3 [8] The variability of basic parameters during evolution.**
The similarity of results for these 4 methods shows the similarity of obtained half-chaos, mainly its evolutionary stability, despite the differences in the way of obtaining. In a-c only met5c and met7ea are shown.
**a** - Stability of parameter *q* (degree of order of the system, the contents of the left peak in Fig.2) shows lack of moving towards the chaos during the evolution - accepting permanent changes which give small changes in the functioning (in the range of left peak, additionally excluded global attractors less than 7, and in the *M20* of met5-7 also smaller than the already obtained).
**b** - The average time of five latest explosions to the chaos (see also Fig.9a,b) does not grow in spite of the above indicated conditions on attractor's length. In the chaotic networks such explosions (see Fig.10) happen almost until the not yet exploded processes exist.
**c** - The average size of local clusters (in met8 they are not checked) and the ice. It makes sense for in-ice-modularity, so not for the met6 where a single local cluster covers the whole network ($N$ = 400). In met7e network *sf* has a specific derogation. A mechanism of it has not been elucidated (see also Fig.1, wider recognition in (Gecow, **2016a**)).
**d** - The average number of global clusters. In the met7 it also stabilizes from the *M7*. In the initial set of initiation (*J*), still without accumulation, it is sometimes even greater than the number generated in-ice-modules, which shows that few so defined clusters may arise within one constructed in-ice-module.
**e** - The average number of local clusters.

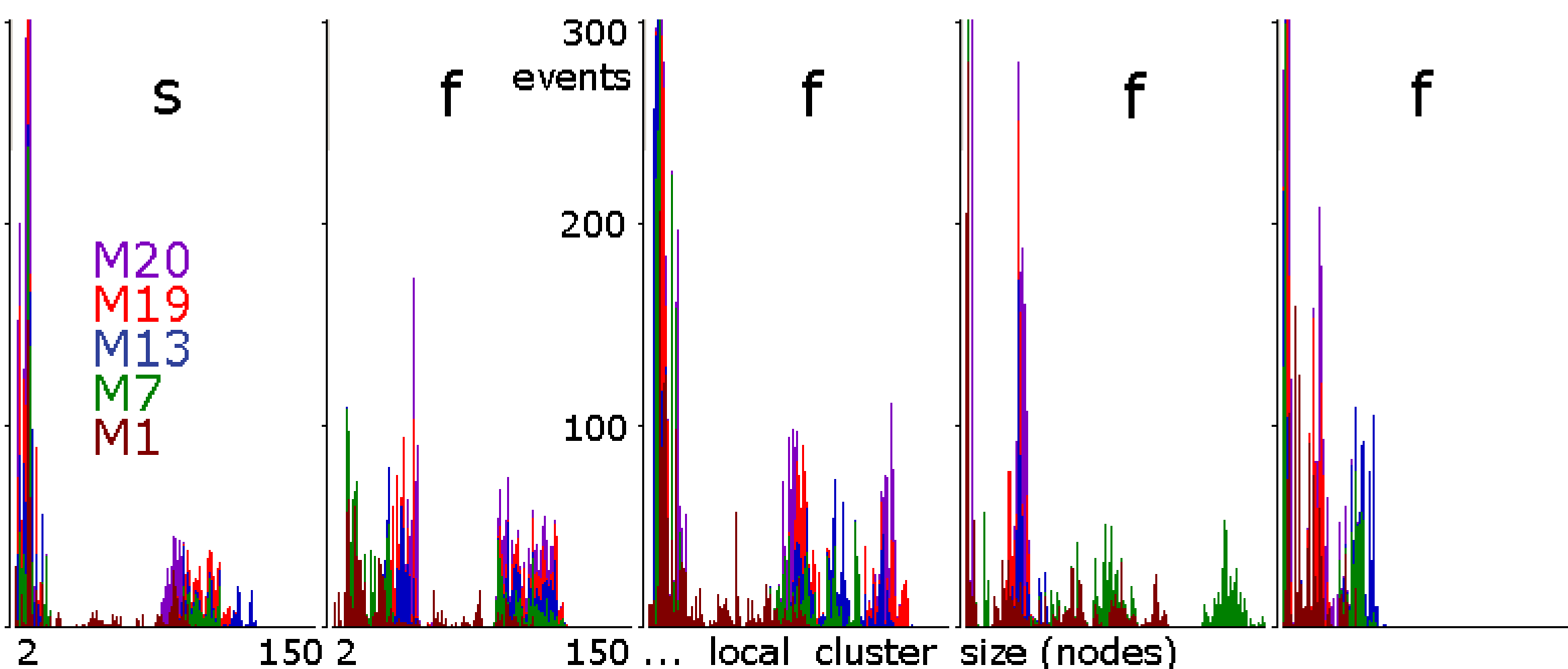

**Fig.4 [2] Dynamical size distribution of local clusters and their stability through evolution met7eb.**
Distributions at end of *M20* of the size of local clusters in the range of up to 150 (of $N$ = 400) nodes collected only in indicated sets. It should be analyzed on the greatly enlarged picture in pixels – one pixel up means one event, to right – one node more in the cluster. Dynamically observed increases are significantly more uneven than this is due to randomness, it can be assessed painstakingly analyzing the size of the growth of a specific color assigned to the particular set *M*, but it does not reflect the image of a dynamic inside the set. This non-uniformity is associated with the presence of different in-ice-modules also changing during the accumulation. Such results are practically identical in met7ea for *er* and *ss*, only *sf* clusters are there typically larger. In the *er,* larger local clusters are very rare. Presented image, especially in the dynamical form in part reproduced through colors, is a strong, eye-argument for the existence and functioning of in-ice-modules. As can be seen, in-ice-modules may even be quite large

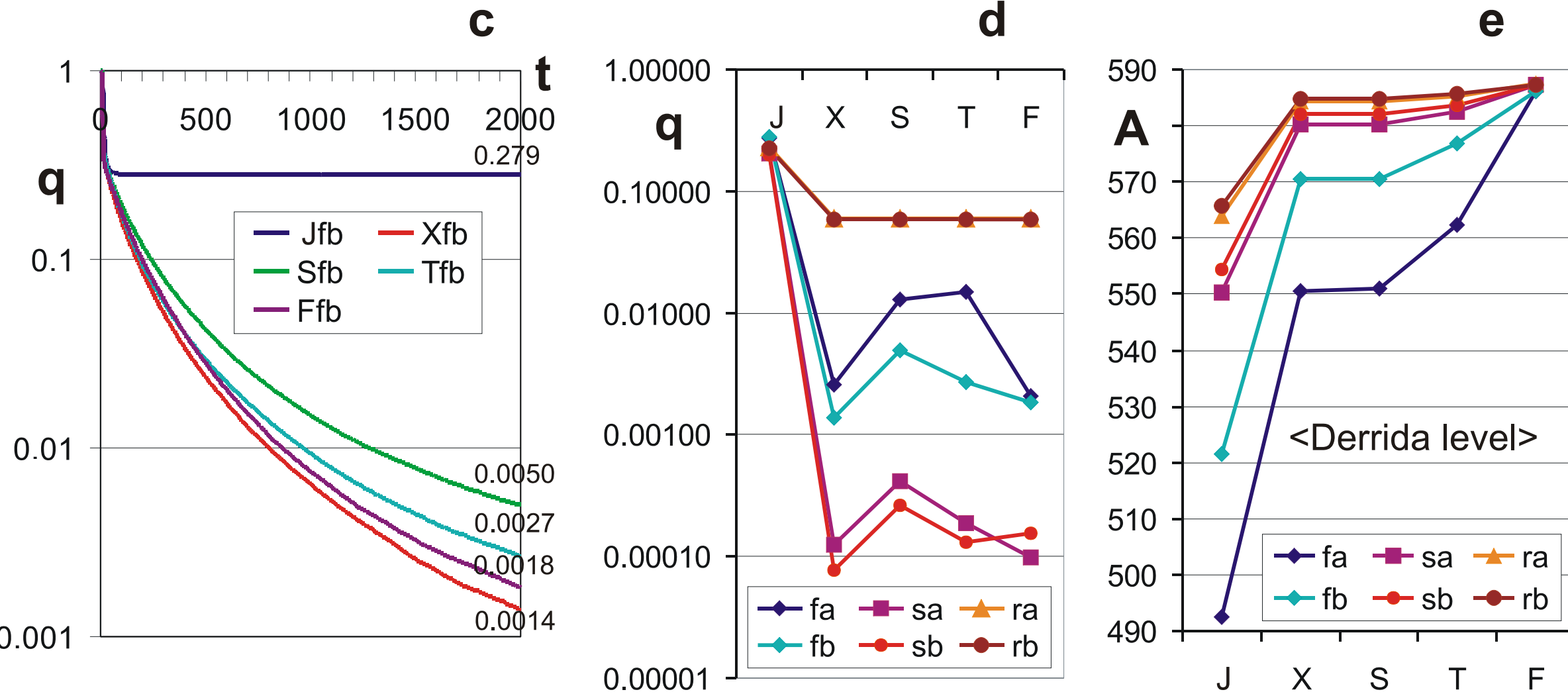

**Fig.5 [10] The difference between half-chaos and chaos in met7a and b.**
Experiments met7a and b (without evolution) were supposed to deeper and more accurately demonstrate the distinctiveness of the achieved half-chaotic state and chaos. In comparison to studying the evolution, an elevated $N = 800$ and $tmx = 2000$ were used. Variant b, over the conditions used in variant a, is forcing small attractors in in-ice-modules, and limitations: local attractor ≤100 and global attractor >200 also a shift to the latest start of local attractor <500. Experiment
*J* - immediately after generation in-ice-modularity (600 networks), and after *J* further experiments *X, S, T, F* (300 networks).
*X* - after acceptance of one chaotic change,
*S* - after changing the node states to be random,
*T* - the shift of functions to other nodes,
*F* - after a generation of random functions for nodes. Despite the lacking possibility of the meaningful designation of measurement errors, the reproducibility of the results and the radical behavior otherness of *J* experiment clearly shows that the obtained state strongly differs from chaos.
**a,b** - Probability of time of explosion to chaos for met7b. This aspect is shown in the graphs of *A(t)* shown in Figs.6, 7 where late explosions resemble the image to the chaotic and increase the uncertainty of the appropriate selection of *tmx*.
**a** - *J* and *X* for network *sf, ss,* and *er*. For *J* the probability smoothly decreases with time increasing, for *X* appears the collapse near $t = 22$ and the transition to a much slower decline associated with the presence of chaotic explosion after the secondary initiations. None of the collapses for the *J* results from the completion of the first round of short local attractor. After this moment there is no explosion as a result of secondary initiation inside the in-ice-module, which would be happened in the new circumstances. This mechanism is an approximation since initiations are also held in the icy walls between in-ice-modules, but there damage spreads more difficult, and after penetration into in-ice-module already subjects to the indicated mechanism. There was a clear difference in the behavior of the tested types of networks - *sf* has later explosions, in this aspect it is the most similar to the chaos; *er* has the least of late explosions.

**b** - *J, X, S, T, F* for network *sf*. Apart from the half-chaotic *J*, the remaining chaotic *X, S, T, F* practically overlap. *X* protrudes somewhat from below, and the *S* and *T* – from above. Very late explosions also occur in half-chaos, but they are rare. These are usually cases of especially large global attractors, sometimes not at all found in the range of *tmx*, furthermore, most initiations appear in the ice between in-ice-modules, where damage normally builds up slowly.

**c** - Average *q(t)* for *fb* (network *sf* in met7b) in experiments *J, X, S, T, F*. Half-chaos in the *J* is clearly different and quickly stabilizes *q*, but *X, S, T, F* drop up to *tmx* and probably further and are a little bit different. In this measurement the difference may be within a measurement error, which is practically impossible to determine due to the multiplicity of factors, but in d at least the *S* and *T* seem to consistently differ from the *X* and *F*. Reviewing diagrams *A(t)* as in Fig.6b similarity is noted in the range of *X, S,* and *F*, but in the case of *T* there are frequent derogation of different nature, particularly for *fa*, where the result is strongly disturbed for a few special cases.

**d** - Average *q* for all the tested types of networks (*sf, ss, er*) and models (a, b) in all the five experiments *J, X, S, T, F*. Network *er* in chaotic cases hides differences due to presence $k = 0$. See also the discussion of differences in the description c above.

**e** - Average position for the right peak of chaotic Derrida balance. Particularly large deviation for the *Jfa* and *Jfb* is shown in more detail in Fig.2 and Fig.3c. *X, S,* and *T* behave here the nature of the derogation and the statistical derogation from the randomness of functions, which suggests such a source of visible here differences and determines the magnitude of the impact of non-randomness of functions on the results. *X* and *S* retain a correlation of non-randomness of functions with node place in the structure of the network, which *T* breaks.

## 3.4 Controlled design of the system with short-attractor

Point attractor, as extremely short, gave sought half-chaos. However, extreme is specific and in the evolution (met5) half-chaos was maintained even when attractor was not found in the range of *tmx* (Fig.7c). It should be checked whether the alone condition of a short attractor, but significantly greater than 1, is sufficient. For that, simulations **met6** causing in the random system a global attractor (of the whole network) = 21 was performed. From $t = 21$ for the unused input states of the node, the function value was changed to state 20 steps backward. **We obtained the evolutionarily stable half-chaos** even with a high *q* (Fig.8) for the same parameters and rules of the evolution simulation as in the met5. **The primary difference is the shape of the resulting left peak** (of small changes) in the distribution of damage size - there are practically only changes of a magnitude $A = 0$, but $A = 1$ and $A = 2$ are present in negligible amounts (Fig.2). This means that practically there are no changes in the functioning and in spite of the acceptance of permanent changes in the functions of nodes, nothing is changed. Such a process **is not suitable for modeling of adaptive biological evolution**, only for neutral evolution. A total **lack of in-ice-modules** was found, but the classic modules are present like in met5. In half-chaos based on in-ice-modularity as in the met5, the peak of a small damage contains a significant amount of change in the range $A = 1$ to 4, and also larger changes occur markedly frequent (Fig.2). In-ice-modularity in met5 explains achieved stability for the larger global attractors - they are assembled of small local attractors (in in-ice-modules), but this solution was checked in met7.

## 3.5 Controlled design of the in-ice-modular system

To determine the sufficiency of the in-ice-modular state to obtain stable half-chaos, we have attempted to controlled create it without booting from the point attractor (**met7**). Networks *sf, ss,* and *er*, $s,K = 4,3$ was studied. First, a network of *N* nodes and their states are randomly generated (dependently on network type). Next, analyzing of the node connections, a collection of 'in-ice-modules' was created and everyone node was assigned to an in-ice-module or separating them ice. Node created new in-ice-module when none of its link (input and output) was connected to a node belonging to an already existing in-ice-module. When it was connected to nodes belonging to only one in-ice-module, it was assigned to this in-ice-module. When it was connected to the nodes belonging to several in-ice-modules or if the limit of in-ice-modules (= 10) or the size of the in-ice-module (= 100 nodes for $N = 800$, 25 nodes for the study of evolution) was exhausted, the node was assigned to the ice.

Next, a trajectory was calculated by appropriately functions selecting. For the current input state, if it was not previously defined, nodes of ice get the value of the function equal to 0, but nodes belonging to in-ice-modules – random value.

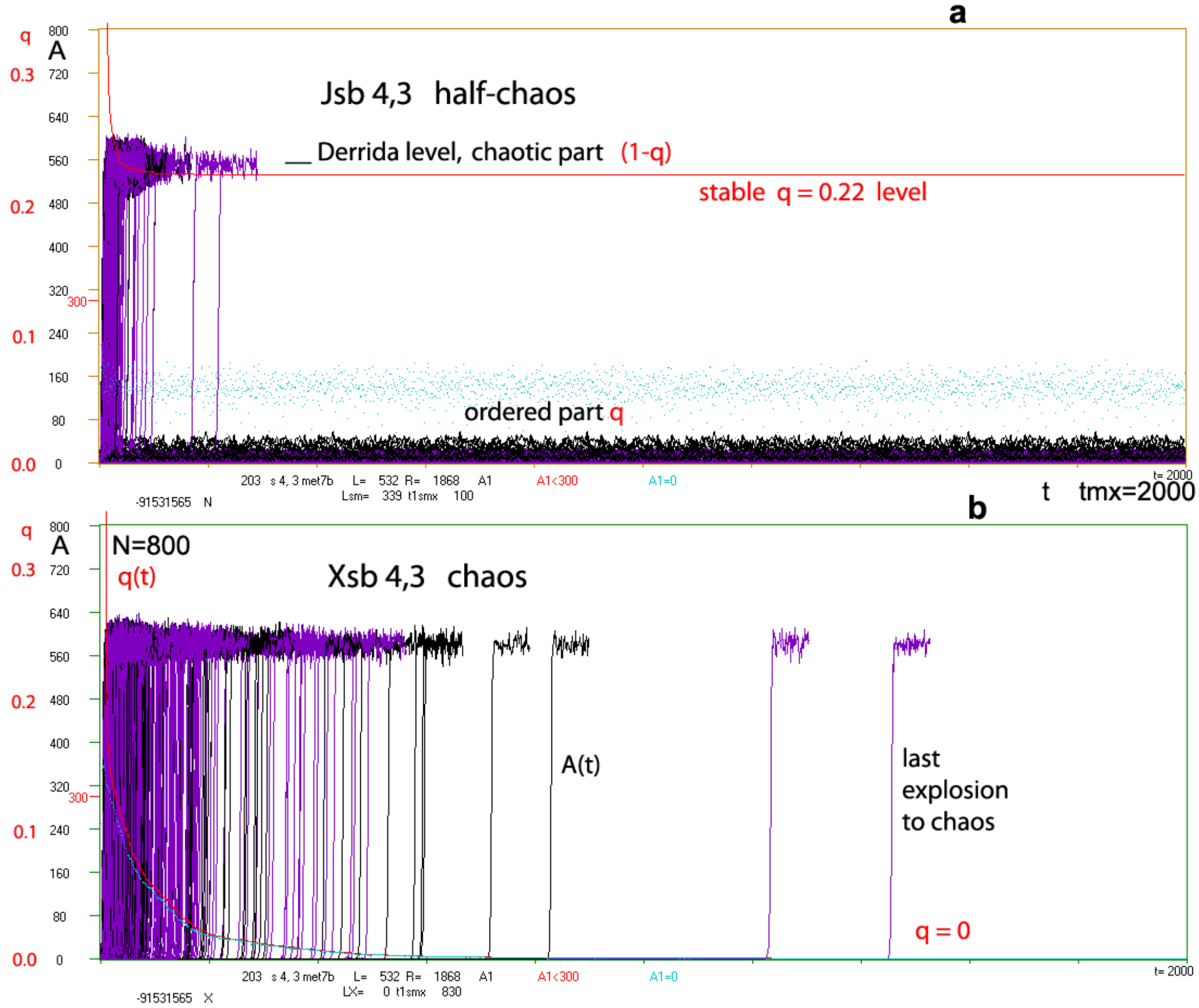

**Fig.6 [3] Half-chaos and chaos in the presentation of *A(t)* for a full set of initiations on the example of met7b *J* and *X* for network *ss*.**

This is a presentation observed dynamically during a simulation on the screen pixels. The details should be watched in enough magnification. In met7b $N = 800$, $tmx = 2000$ was used. A rectangle has the dimension of 400*1000 pixels, so on each axis, one pixel shows 2 values. In Fig.7, for which this figure is a description of form, $N = 400$ and $tmx = 1000$ is used, so the there unit on the axes corresponds to a pixel. The vertical axis is originally scaled in the *A* - number of the nodes states different than in the pattern. The horizontal axis is the number of steps *t* of simulation of network functioning. After each initiation by small permanent change, the state *A(t)* was drawn with a continuous line on the screen after every step of the calculation. In case of initiation of a node in the in-ice-module black color is used and for initiation in the walls between in-ice-modules - purple. In met5 shown in Fig.7, this distinction was not known and always black was used. To optimize the simulation a counting after 70 steps from the explosion to chaos (crossing over the threshold, here = 300, marked in red on the left) was stopped - there the process has no chance to return.

As can be seen, the transition to chaos in the vicinity Derrida balance is not slow, but rapid in several to over a dozen steps, where A increases drastically, so - "explosion." After deflection from a small value to say - $A = 80$ no longer the returns happened (as checked without optimization, see (Gecow, **2016a**)).

After the end of initiation set, the red curve *q(t)* was added to the figure. In met7 it is originally scaled by the A as the number of initiation, which does not exceed the threshold = 300, but there are $3N = 2400$ of initiations. In met5 in a Fig.7 *q(t)* is divided by the number = 3 of initiation in node, so that $q = 1$ for $A = N$. The red description in the left has been added for readability and here *q(t)* is the share of processes that in the time *t* did not pass the threshold.

**a** – Half-chaos, experiment *J* for network *ss,* model b. There were 600 of such simulations for each type of networks *sf, ss, er* and models a and b of met7. The red curve *q(t)* quickly stabilizes at a high level $q = 0.22$. In the lower part of the graph, many trajectories are visible (there are L = 532 of 2400) that a little over $t = 200$ no longer explode. So R = 1868 processes from the very beginning went to chaos - a Derrida balance.

**b** - Chaos on the example of experiment *X* performed immediately after the measurement of the *J* illustrated above in the a. There were 300 of such simulations for each type of networks *sf, ss, er* and models a and b of met7 and for each experiment of *X, S, T, F*. Here, *q(t)* is steadily decreased until all the processes are not 'exploded'. At the end there is exact LX = 0 of them, means $q = 0$.

Blue points describe the number of processes that currently have $A = 0$, i.e., damage fade out, but for the *X* the secondary initiations lead to their explosion.

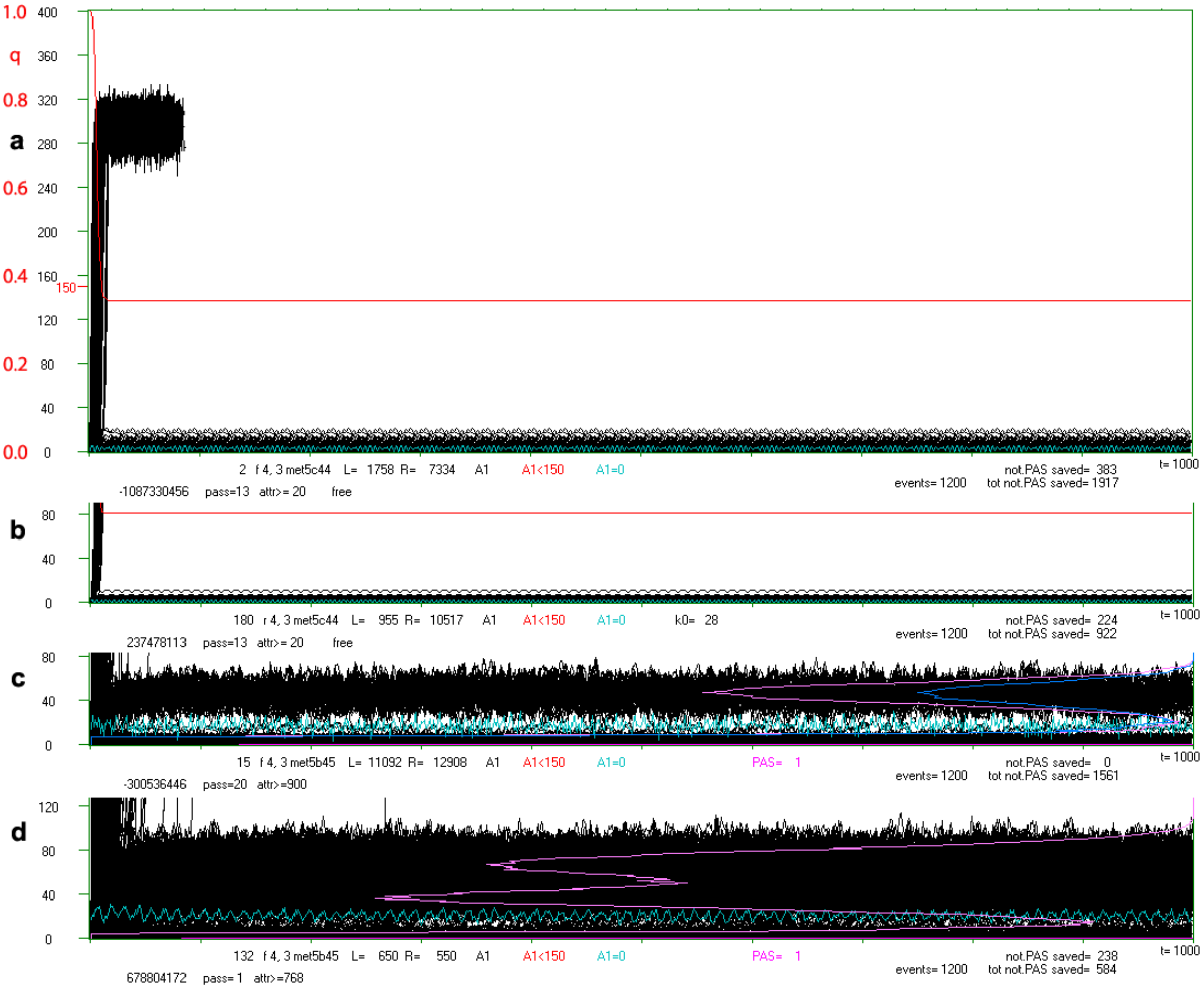

**Fig.7 [5] Simulations met5 (changes accumulation) in the presentation of *A(t)*.**
Except for red description *q* on the left, each drawing was created dynamically on the screen during the simulation of one full set of initiation without blocking of reverse initial changes. It is accurate to the pixel. Description of the presentation elements in Fig.6.
**a** - Full typical image for the *M13* met5c (met5 in other figures, model c from met4), network *sf*. Almost an immediate end of the explosions to the chaos can be seen. At the top - the state of chaos in the Derrida balance (short due to optimization by interrupting the counting after 70 steps, as in Fig.6). At the bottom - a repeating pattern in accordance with the global attractor marked on the top frame (pattern network state as in *tmx* before the first initiation of the set). Here L and R under the lower frame is the sum from the beginning of the evolution simulation of this network. In this set 383 of initial changes were accumulated of 1200 tested, but accepted changes defining *q* (not exceeding the threshold = 150) were a little bit more (with global attractor <7).
**b** - Typical image of network *er* simulation in met5c. The upper part of the almost identical to a is cut. The level of *q(t)* is lower, the belt at the bottom - clearly thinner, the time of the latest explosion to chaos - shorter.
**c, d** - The lower part of the image for met5b (with minimal regulation). Here the level of *q(t)* was much higher than in a. In the model b, the width of the lower belt is greater due to the possibility of regulation. Simulations slightly different model than in Fig.10d - here without blocking of reverse changes, but with the condition non-decreasing of global attractor and accumulation of changes not less than $A = 3$, the shift of beginning = 2, but not 50. In these simulations, a distribution of damage size for ordered cases ($A < 150$) was studied on the section from $t = 600$ to *tmx* for a given set of initiations (purple curve on the right frame) and the sum of the sets in the final set *M20* (blue curve in c). It is one of several ways to look for proof of the in-ice-modules existence. As can be seen, in both (c,d) shown cases in these distributions the significant peaks are visible. They indicate an existence of one (in the c *M20*) or two (in d *M1*) hypothetical in-ice-modules. Under the scope of these peaks, there is a clear gap in the minimum of distribution. An interpretation of these peaks can vary, they are not proof of the in-ice-modules existence, which was shown later watching nodes states repeating, but they are a strong premise.

The $q$ level here is high: in c $q = 0.46$ and in d $q = 0.55$. In c the attractor was not found at the beginning of the set (attr ≥ 900), and because it could not decrease, no one accumulation happened (not.PAS saved = 0). It does not mean, however, that there is no here acceptable ($A$ <150) cases (there are 220), which indicates $q$ and wide black belt below the $A = 150$

A number of additional conditions and adjustments was applied, documentation (Gecow, **2016a**) contain a full description, their details are not important here. **Initially, short attractor was forced in each in-ice-module** and using this assumption basic investigations were made: (**b**) – of the in-ice-modularity state (series with **$N$ = 800 and *tmx* = 2000 without evolution** roughly corresponding to the met4) and (**eb**) - **the evolution** as in the met5 and met6 (series with **$N$ = 400, tmx = 1000**). In the end, the necessity of **this assumption** was verified and surprisingly it **occurs unnecessary**. So the two most important research without the forcing of the short attractor in in-ice-modules were repeated (called **a** and **ea** - as logically simpler).

Examination (***J***) of the in-ice-modularity with **$N$ = 800** mainly relied on checking the $q$ and the distributions of damage size. In the versions b, we demanded the global attractor to be greater than 200 when the local attractor could not exceed 100 - the result was in line with the tested vision which explains the admissibility of larger global attractors. **In both versions (a and b) it was verified that the statistical properties of non-randomly selected functions are not responsible for the increase of stability**, namely - how such a system behaves after: the acceptance of one large change (***X***), randomly changing of node states (***S***), moving the functions to other nodes (***T***), and the random generation of new functions (***F***). In the experiments *X, S, T* functions retained their statistics. In all these experiments chaos yielded (like *X* in Figs.8, 6b), but it systematically slightly differed from the full version of chaos *F* (Fig.5).

Comparing with the met5, particular for network sf, both peaks of the distribution of damage size have been a little bit changed (Fig.2). Also in distributions of the ice size and the local clusters size the blur arise what caused a marked decreasing of average ice and increasing average size of local clusters (Fig.3c). This shows getting a **slightly different state of in-ice-modularity**. Like in the met5 and met6, system parameters stabilize from the *M7* and **the small change as a condition of acceptance is sufficient to any long maintain of half-chaos in the version of such the in-ice-modularity.**

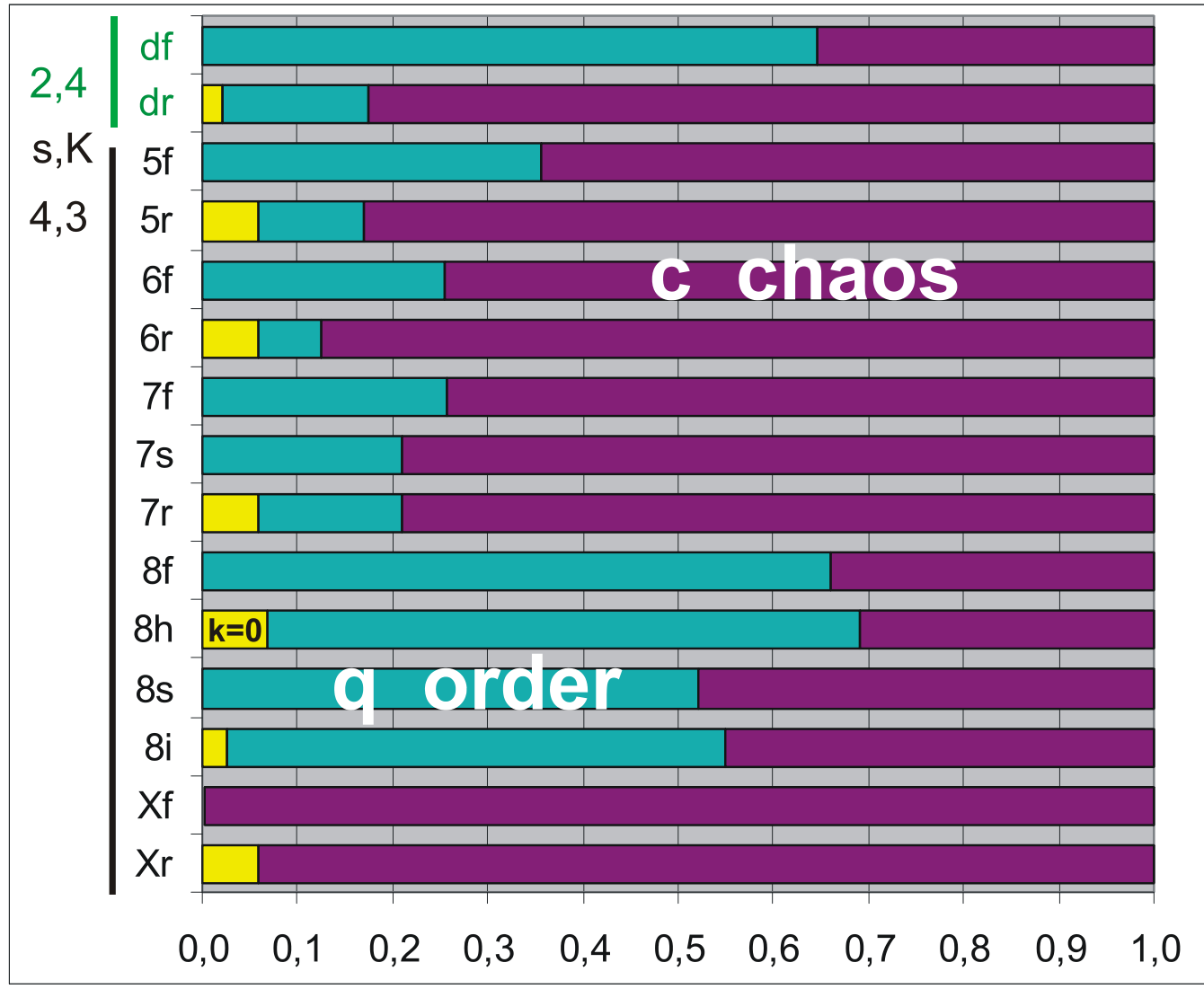

**Fig.8** [9] **Half-chaos – fractions of ordered events ($q$) and chaotic ($c = 1$-$q$).**
Experiments described as in Fig.2. In the range of $q$, an order resulting from the absence of output in some nodes ($k = 0$) in the network *er, sh* and *si* is isolated as yellow. All results presented here concern only the effects of length limitation of global attractors (‘6’ - met6) or length limitation of local attractors through in-ice-modularity. For ‘8’ local attractors are not detected, but the level of ice (Fig.3c) shows, that local clusters cannot be large. For ‘d’ and ‘*X*’ (met7a) there is no evolution, the results concern the network immediately after generation of half-chaos, but for ‘*X*’ also after acceptance of one chaotic change, which gives a typical chaos (see also Fig.3f). In the remaining methods (‘5’, ‘6’, ‘7’ and ‘8’) result is a sum of the results of 4 stable complete *M*, as in Fig.2, (see Fig.3). Except ‘d’ where $s,K = 2,4$, in remain cases $s,K = 4,3$. See also Fig.10d.

### 3.6 Growing half-chaotic networks

Much more complicated and stronger is the **disturbation of a system through adding or removing a node (met8)** (Gecow, **2017a**). There are problems with the comparison to the undisturbed system and the interpretation of secondary initiation. These simulations **start from a small system (*N* = 50) with a point attractor**. The network grows in 5 successive stages *M* by 100 nodes and reached *N* = 550 at the end. The overall picture was very close to met5 and met7. Also, **half-chaos** (Fig.2 right, Figs.3a,b, 8) **with evolutionary stability and stable presence of large ice share are obtained** (Fig.3c). It suggests similarity of mechanisms of increased stability to in-ice-modularity. In this case, the network grows by evolving under the control of a small change. The gap between the right and left peaks is not so empty here (Fig.2), probably because adding or removing a node is not a very small disturbation.

# 4 Supports for stability

### 4.1 More of negative feedbacks in a random system, function narrowing

It is generally believed that the stability of the various systems results from homeostasis based on regulation by negative feedbacks. Kauffman pointed instead to the property of the ordered phase (order for free) (Kauffman, **1996**) as the most important reason, but for it extremely small *K* should be expected. **The regulatory feedbacks** are generally considered the basis for the stability of living entities and their **concentration is considered to be significantly increased** in relation to the random one. **However, the complex structure of the feedbacks for this statistical surplus has been replaced in the Kauffman model by their proper effect (ice)** and it remains only in random share. So much simplified model is not able to give a proper statistical picture of a system failure and conclusions for a stability mechanism can (and seem) significantly differ from reality.

This doubt was the main reason for undertaking the research, which initially aimed to strong raise the share of regulatory mechanisms.

In the presented study **we transform part of the feedbacks in random structure into negative feedback**. It is done by changing the random function when the state on the inputs was not used yet. It was the first method (**met1**) of correction of a random chaotic system. The similar, stronger **met2** has iterative change the pattern. Network *s,K* = 2,4 and 4,3 were investigated. Fig.1c suggests that Derrida chaotic balance is achieved even before the 15-th time step. **Initial research for *tmx* = 60 steps yielded very promising results** (Fig.9a) - *q* was significantly increased (especially for *s,K* = 4,3), the distribution of damage size already contained two peaks separated by a gap. A large part of this effect (especially for *s,K* = 2,4) was the result of deviation from the randomness of node functions (function narrowing), which also may be included (Kauffman *et al.*, **2004**) to evolution tools. But it turned out (Fig.9) that obtained in met2 stability of ***q* usually significantly decreases with the elongation of *tmx*,** practically disappears already for *tmx* = 1000, only in the case of Boolean networks *sf* 2,4 this method could be considered to be effective to achieve half-chaos (not tested for evolutionary stability). As it can be seen in Fig.9bc, *tmx* = 20,000 was used. Simulation series contains 700 nets for *s*=2 and 350 nets for *s*=4.

These studies demonstrated a **high range of results dependence on the network type** - the network *sf* is more ordered (Iguchi, Kinoshita, Yamada, **2007**); network *ss* and *er* are more chaotic, similar to the reaction, but *er* has part *k* = 0 (Figs.8, 9, 10) obstructing observation. **The parameters *s,K* = 2,4 and 4,3 also give a very different picture**. The simulation allowed for a deeper look at the process and its determinants, which **pointed to the short attractor** (ch.3.1).

### 4.2 Modularity

**It seemed that the most natural way to get short attractors is modularity. In met1 and 2 no modular effects were observed,** although modules exist in practically every network. It was assumed that in a random network the modules are too 'weak', then it was pre-checked, what stronger modularity gives for stability (**met3**). Here it turned out that sufficiently small spontaneous attractors can be expected only in so small modules that the consideration a state of chaos in them losing meaning. Consideration of chaos in the modules network has been postponed.

In the study, the network has ***N* = 400 nodes. It was assembled of *N2* = 50 modules, each of *N1* = 8 nodes**. Connectivity *K1* between nodes inside modules *K1* = *K2* connectivity between modules. The rule of connection is taken like in type *er*. Simulation series consists of 100 nets.

**The modularity also gave raise *q*** (Fig.9c), **especially when met2**, which increased the share of negative feedback, **is used at the same time**, however, evolutionary stability was not checked. In the distribution of damage size, the typical for the half-chaos radical **gap between peaks was not observed, only the clear minimum**. An increase of $q$ in the experiment met3+met2 with $s,K$ = 2,4, almost entirely resulted from non-randomness of functions (**function narrowing**). Both of these methods and their associated factors (such as function narrowing) belong to the most important methods of producing desired stability by biological evolution, but in both the short attractor is an important factor.

As was described in ch.3.3, classic modules cooperate with in-ice-modules. The theme of classic modularity and its role in system stability was here recognized only provisionally and requires much deeper research. However, it is one more source of modularity, than was found in (Altenberg 2005), where the role of modularity is studied in depth in evolution.

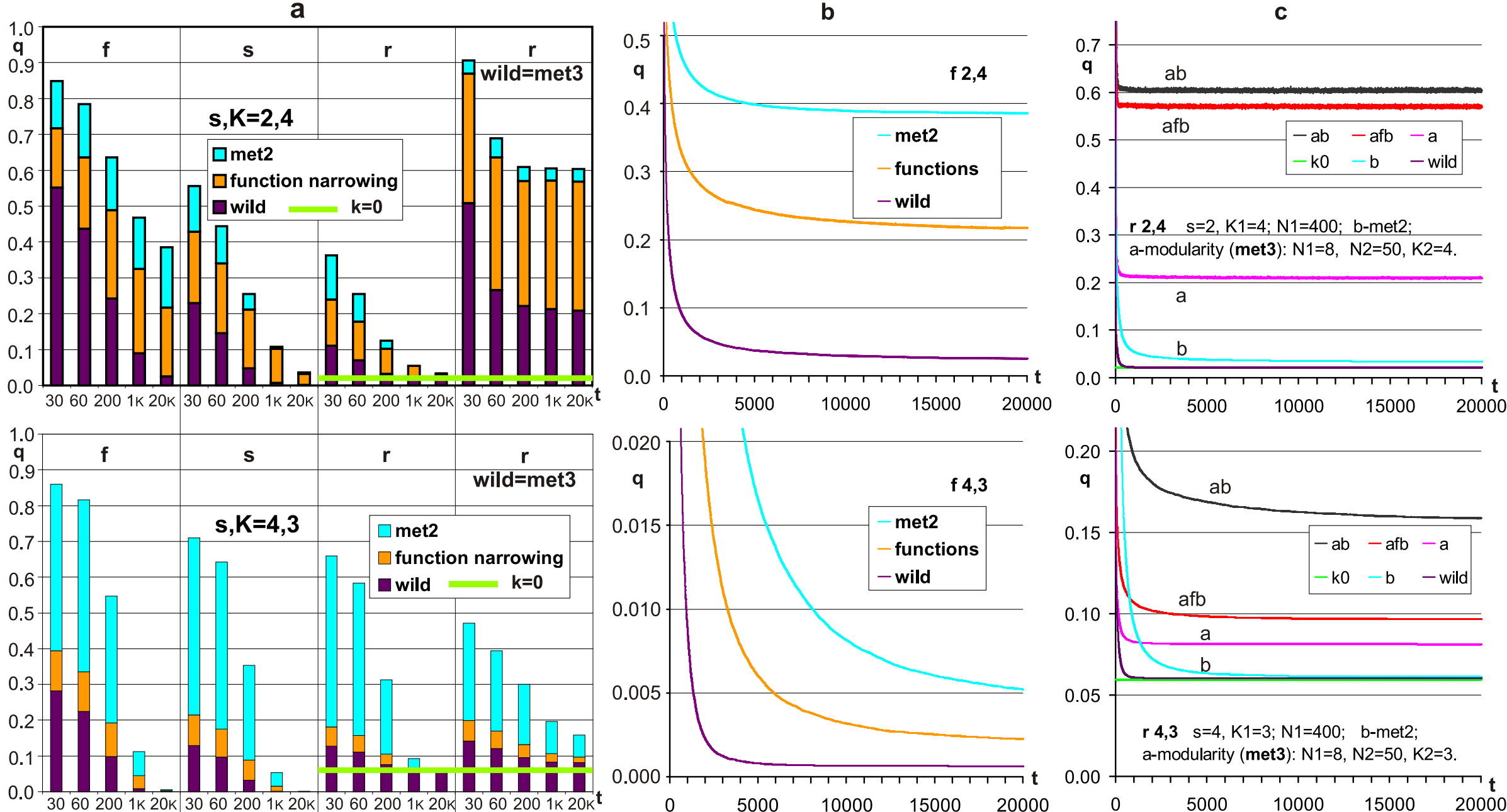

**Fig.9 [4] Ordered fraction ($q$) as a function of time ($t$) after raising the share of negative feedbacks (met2) and the classic modularity (met3).**

The upper row of all part - $s,K$ = 2,4 (Boolean network), lower - $s,K$ = 4,3.

**a** – For some moments $t$ the shares of mechanisms: wild - without interference met2; function narrowing as a side effect of the method; the increased participation of negative feedback by met2. For network *er,* the level of q resulting from participation $k = 0$ (nodes without outputs) is indicated by the green line. In the right column as a wild the modular system resulting from met3 is used, further described in (c) as a curve a. The type of networks *sf, ss, er* is described by a second letter. As can be seen, the results for the simulation parameters $s,K$ = 2,4 and 4,3, and network types, differ significantly. For $s,K$ = 2,4 the function narrowing is of utmost importance to increase $q$, but for $s,K$ = 4,3 the importance of feedback turns out to be essential. For small $t$ the effect of increase $q$ is significant. From these data it can be suspected to achieve half-chaos for: *sf* 2,4 - the result of functions narrowing and increase of the share of regulatory feedback, and for the assembly of modularity met3 with met2 using nets *er* - for 2,4 mainly due to the functions narrowing, but for 4,3 due to the met2. In the remain 5 presented cases the effect practically disappears already for $tmx$ = 1000, the use of it by living entities require very rapid multiplication in comparison to the transformation of the construction and metabolism, which seems unattainable. Here evolutionary stability (included in the definition of half-chaos in the result of further studies restricting fundamental factors to a short attractor Figs.2, 3, 8) was not examined. The degree of entry into the plateau can be better assessed in **b** and **c**. The network ss gives a similar effect to the network *er*, but without the confounding effect of $k = 0$.

**b** – Net *sf* 4,3 (350 nets) not reached a plateau even at $t$ = 20,000, where $q$ is negligible, but *sf* 2,4 (700 nets) is almost on plateau $q$ at $t$ = 5000, and this level is high (compare Fig.10d).

**c** – The result of modularity (met3) and assembling it with met2. Result of met2 for network *er* is added, such as in **b**, omitting, however, the share of function narrowing enough presented in **a**. It can be seen that the wild system (without forced modularity, 700 nets for $s$ = 2, 350 nets for $s$ = 4) of network *er* very quickly descends to the level of $q$ resulting only from $k = 0$. Also curve b - the result of the met2 quickly closer to that level, which can also be seen in **a**. Forced modularity (curve a, 100 nets) gives a clear stable increase of $q$, and met2 help it (curve ab) to radically increase $q$, but for $s,K$ = 4,3 appears to fall within the plateau above $t$ = 20,000. For $s,K$ = 2,4, almost all large and stable met2 effect results from the function narrowing only (curve afb). The network has $N$ = 400 nodes assembled of $N2$ = 50 modules each of $N1$ = 8 nodes.

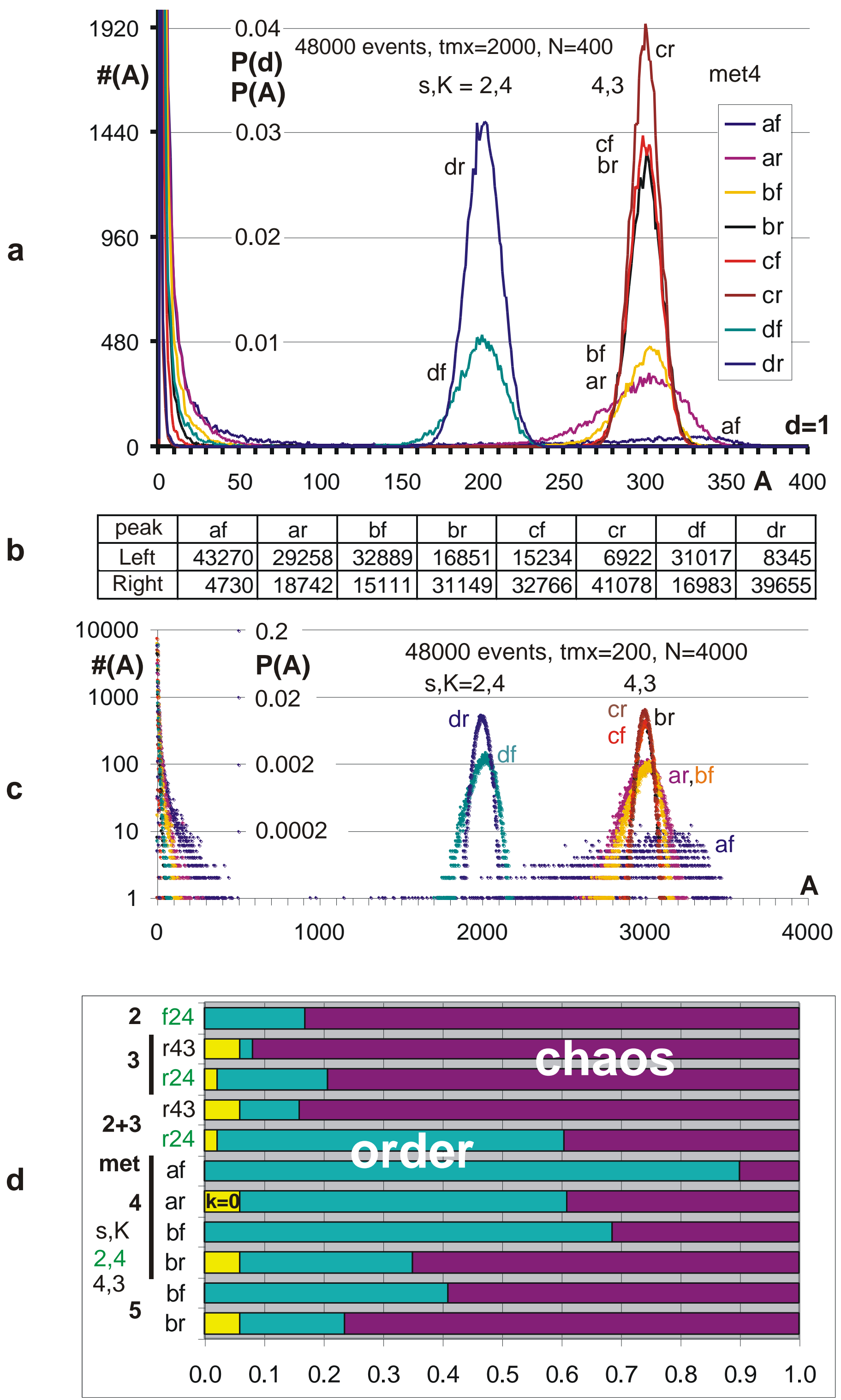

| peak | af | ar | bf | br | cf | cr | df | dr |
|---|---|---|---|---|---|---|---|---|
| Left | 43270 | 29258 | 32889 | 16851 | 15234 | 6922 | 31017 | 8345 |
| Right | 4730 | 18742 | 15111 | 31149 | 32766 | 41078 | 16983 | 39655 |

**Fig.10** [6] **Increasing regulation or another factor - the point attractor. The primary result of the met4.** In the met4 removing a presumed cause of the poor performance of the met2, we start with the non-random system with extremely short attractor – a point attractor: initially, all states are set to 0 and f(0) = 0 (f – node function). The models were tested in the sequence a, b, c ($s,K$ = 4,3) and d ($s,K$ = 2,4) starting from strong regulation and ending with the lack of regulation in the models c and d. Care was taken that each signal has the same probability in the function of each node.

**Model a** contains a negative feedback with a positive (1) and negative (3) deflection from equilibrium (0) in each of the three input signals. It contains also the leaving of homeostasis into the area of randomness (when deflection is too great or one of the input signals = 2, then the node function is defined randomly). A more exact description of this formula can be found in the text of ch.4.3 and is available in (Gecow, **2016a**).

**Model b** has a minimal regulation: the condition of the point attractor f(0,0,0) = 0 is supplemented only by condition f(0,0,1) = f(0,1,0) = f(1,0,0) = 0 that there is no in **model c**.

**Model d** of Boolean network ($s,K$ = 2,4) has only condition f (0,0,0,0) = 0 similar to model c.

Each model is simulated for three combinations of $N,tmx$ = 400,200; 400,2000; 4000,200 for networks *sf* and *er,* so as to always number of initiations was 48,000 in the series. The threshold of small change for $N$ = 400 was set to 100, and for $N$ = 4000 to 800. Each initiation by definition of met4 is made for node state = 0 and for input state = (0,0,0). So only in the model c 3 other function values may be used for initiation. For model a the only one value 2 remains, for b only two values: 2 and 3, which are new states of a node without the mandatory fade out of damage at the destination.

**a,c** - The counts $\#(A)$ of processes ending in *tmx* with value $A$ (changed states of nodes in *tmx*) are shown. Also, the scale of the $P(A)$ or $P(d)$ are added. The results showed here in the linear plot a ($N$ = 400) for models c and d are also in Fig.2 in log scale. The series showed in c contains 10 times fewer networks, which gave peaks much narrower (in damage $d$ scale instead of $A$) than in a. The right peak for models b, a is becoming smaller due to increased regulation, which is reflected in the diagram d as less participation of chaos. Place of the right peak in a and c are well designated by Derrida balance (Fig.1d) (different for $s$ = 2 and $s$ = 4), which is the property of a mature chaos.

**b** - The table of results $\#(A)$ for *tmx* = 200 for the same networks as in a for which *tmx* = 2000. The counts differ only for af by 140 and for df by 2 (less for left peak).

**d** - A complementary for Fig.8 juxtaposition of a fraction of ordered cases ($q$) and chaotic cases ($1$-$q$) for minor experiments discussed in the article. While Fig.8 lists only the study of impact of small attractor, it is here - the impact of increasing the share of regulation in met2 (only *sf* 2,4 can be considered in met2 as entry into half-chaos, see Fig.9a,b); of modularity in met3; assembling of met3 and met2 (Fig.9); assembling of point attractor and regulations in met4ab and met5b. Among them only met5b examined the evolutionary stability included in the definition of half-chaos. As can be seen, the assembling is more effective than approach alone and should be expected of such a strategy in biological evolution. The case af shows that the way evolution can lead to a state where the half-chaotic system may seem as ordered. Evolution met5b decreased $q$ comparing met4b when met5 (Fig.8) worked in the opposite direction relative to met4c (these are uncertain trends), but the expected strategies of biological evolution its creative aspect is important, not modeled in the presented simulations, too simplified to such a task.

## 4.3 Regulation in system with point attractor

Lack of expected radical effect of regulatory mechanisms in the met2 was found in the system starting from a random network, then we introduced strong regulation in a system with a radically short attractor – point attractor (**met4a**). This time the result was surprisingly strong (Fig.10), so we decreased the regulation to the minimum (**met4b**, see also **met5b**, Fig.10d) and next, regulation was rejected at all (**met4c,d** and later), which showed that the point attractor is sufficient to achieve half-chaos.

In met4 point attractor starts with all node states equal 0. It was not permitted in met8, where states are random. Model **met4c** for $s,K$ = 4,3 used later in met5 is defined as f(0,0,0) = 0. **Model d** for $s,K$ = 2,4 – as f(0,0,0,0) = 0. They are based only point attractor, without regulation. **Model b** with minimal regulation ($s,K$ = 4,3) also used later in met5, has in addition to c also: f(0,0,1) = f(0,1,0) = f(1,0,0) = 0. For signal value 1 interpretation was taken: deviation from proper state '0', but still in the range of homeostasis. For **model a** ($s,K$ = 4,3) also f(0,0,0) = 0, but description is much more complicated. Here direction of deviation in homeostasis range: 1 – positive; 3 – negative. The deviation of one of 3 input signals gives 0. The function also gives 0 if 2 signals are deviated, but in the opposite direction and third is 0. If 2 signals deviate in the same direction but third is 0 or 3 signals deviate, but they are not equal, then function result is deviated, i.e., is 1 or 3. If 3 signals deviate and are equal or at least one is 2, then the result of the base function is 2, but such value for a particular node is converted into random value in the way that share of each function value be equal. Other parameters of simulation in met4 are described in ch.3.3.

**The result of met4a shows how strong may be the effect of the regulation in the half-chaotic system** – right peak almost disappears, that is the probability of entry into chaos as a result of a small system failure (internal cause) is small. **This gives a deceptive picture of the ordered phase** (Shmulevich, Kauffman, Aldana, **2005,** Serra *et al.,* **2007**). There remain external causes, which model of the autonomous network does not take into account from assumption. However, adaptation is to the environment, which can vary and the evolution should be tested using open systems as in (Gecow, **2009**).

# 5 False assumptions of Kauffman's model – Summary

The Kauffman's widely known hypothesis "life on the edge of chaos and order" (Kauffman, **1993**, **1990**), pointed out an important factor in modeling of biological evolution, processes in social organizations, and technical constructions, however, it was based on too simple model, even – on few false assumptions:

**1.** Any network of conditions can be described as Boolean, then it is sufficient to study the Random Boolean Networks (RBNs). Such complex networks are finite, discrete, deterministic, and fully random.

The assumption that the statistical properties of Boolean networks are general is false (Gecow, **2011**). The number *s* of equally probable signal variants should be also considered higher than only two.

**2.** RBNs can be either ordered or chaotic, which is observed and confirmed by the current mathematical theory of chaos.

The current mathematical theory defines chaos by Lyapunov coefficients in **infinite, continuous space**. High sensitivity to initial conditions, leading to maximally different effects for very similar initial conditions is the main characteristic of the chaotic behavior of dynamical systems. Kauffman (**1993**) uses such the term 'chaos' to describe **finite, discrete networks**. **The term 'chaos' is not reserved for just one of those separate areas**. This theory is used for finite discrete networks (e.g. Turnbull *et al.* **2018**), but such a method **is an approximation, which loses a few important phenomena,** e.g., repeating the same argument for a function, the path length to the attractor and attractor length (in steps of a process). Analog of Lyapunov exponent for networks (coefficient of damage propagation (Gecow, **2011**), eq.4.8 in (Serra *et al.,* **2007**) or eq.6.2 in (Aldana, Coppersmith, Kadanoff, **2003a**)) in the case of half-chaos turns out to be misleading.

**Model-forced strong limitations on parameters** are not compatible with estimations from nature (Aldana, Coppersmith, Kadanoff, **2003a,** Luque, Ballesteros, **2004,** Sole, Luque, Kauffman, **2000,** Turnbull *et al.* **2018**). For the evolution of life, the model allows only extreme $K = 2$ (connectivity, *K*—number of node inputs) and $s = 2$. Higher values of *K* or *s* lead to useless chaos.

**3.** Random networks contain all possible networks, then it is not important that **living organisms are not random** in the aspect of stability due to natural selection. Many works have assumed that this stability is explained by natural properties of the ordered system known as "order for free" (Kauffman, **1996**). These are false assumptions. Such a picture was not very consistent with the observed delicacy of living entities, not emphasized of regulatory structures and did not contain a model of death necessary for the Darwinian elimination. Kauffman (**1993**) considered negative feedbacks, but practically (Gecow, **2011**) he left them on a random level.

**In this work, it is experimentally shown that among discrete and finite systems that are not fully random, with parameters *s* and *K* which for fully random system result in chaos, there is a third state of systems** I call **half-chaos**. The not fully random networks where half-chaos is found are obtained considering the specific correlation of parameters which Kauffman simplifying took as random. The analogy to the phase transition is more complex here - it is rather the "superheating".

The particular half-chaotic system exhibits small and large damage. Current theory does not foresee such the possibility, but it is easy to show examples using computer simulation – system with point attractor is half-chaotic. The modeled objects (like living or administrative units, technological processes, and technical constructions) are certainly neither infinite nor continuous. Half-chaotic systems better describe the modeled objects, **freeing modeling from difficult theoretical limitations** (see point 2 above), which until now are the typical basis of many considerations (Nghe *et al.*, **2015,** Serra *et al.,* **2007,** Aldana, Coppersmith, Kadanoff, **2003a,** Kauffman, **1996,** Luque, Ballesteros, **2004,** Sole, Luque, Kauffman, **2000,** Villani *et al.*, **2018**). This opens the door to adequate models with complex networks.

The large gap between small and large damage defines in a natural way a small change, which is very important for interpretation. The peak of great changes (of functioning—damage) well model a death and elimination. After the great change, the system becomes forever simply chaotic, but a small change retains half-chaos and identity of the system, then evolution can go on. This feature as 'evolutionary stability of half-chaos' was included in the half-chaos definition. Half-chaos together with given initializing changeability completed by the multiplication of evolving system resulting from the demand of long evolution offers the full basic Darwinian mechanism.

The Kauffman model is trying to describe living systems and similar ones using several easy to show, and it would appear that the main parameters, the rest of them simplifies assuming their randomness, but natural selection works on all possible parameters, which may be easier and more important for selection and its effect. Indeed, it is difficult to imagine the possibility of the existence of half-chaotic systems from Kauffman point of view. In fact, after the system is drawn, it is either chaotic or ordered (ad. 'observed' in point 2 above) and the set of random systems contains all the possible ones (point 3 above). In the interpretation of the results of this approach, it has not been seen that the statistical absence of intermediate systems does not imply a small number of such systems. There are a lot of half-chaotic systems, but their share is negligible because there are radically more chaotic systems with given parameters (e.g., *K*) - for larger *N* not imaginable many. Model GRN based on RBN is not false, but its assumptions are too simple. Each model is a simplification, for some applications it can be useful, but if it gives a false expectation of important parameter, then some simplifications must be rejected which is the next step of approximation. It cannot be found without a previous step.
Regulatory feedbacks (misinterpreted and practically included only on the random level in the Kauffman model (Gecow, **2011**)) also the classic modularity and narrowing of the function significantly increase the stability, which was noticed, but the main and the new condition is the short attractor. They take over the role of explaining the experience (Shmulevich, Kauffman, Aldana, **2005,** Serra, Villani, Semeria, **2004,** Villani *et al*, **2018**) from "order for free", which in the half-chaos lost importance. The reached a deeper interpretation of Kauffman hypothesis gives a picture much more consistent with the observation and indicates systems more adequate to the modeling of biological evolution. This significantly alters the existing basis of many considerations and probably their conclusions. Likewise, the description of the systems from 'liquid' region (Kauffman, **1990**), where Kauffman saw living objects - "small lakes of activity in the ice" (originally (Kauffman, **1990**): “unfrozen islands”) remains valid for the primary and the most appropriate form of the half-chaos for the evolution - in-ice-modularity discovered in these studies. The base of the in-ice-modularity is an activity of nodes (they change their states) in the ice (where nodes do not change their states), however, in-ice-modularity is supported by classic modularity, which is always present.

# 6 Conclusion

This work examines the systems described by networks that are: autonomous, complex, finite, discreet, directed, functioning, deterministic, and designed as the Kauffman network (Boolean), but with the admission of more than two ($s \geq 2$) equally probable signal variants. The number $K$ of the node inputs in a given network is fixed. Parameters $s$ and $K$ have values ($s = 2$, $K = 4$ or $s = 4$, $K = 3$), which in the case of fully random networks give unambiguous chaos.

Half-chaos is the state of such a system in which small disturbances cause both small and large 'damage' (changes in the system’s functioning) occurring statistically similarly often. As a reminder, in the chaotic system there are only large damages and in the ordered - only very small. The studied half-chaotic systems are not fully random, but they have typical characteristics indicating a full randomness.

The evidence presented in the work indicates that half-chaos is an experimental fact. Its basic mechanism is based on a short attractor, but it is too weak a condition for modeling adaptive evolution. Much more adequate (to describe the purposeful systems) the half-chaos variant depends on in-ice-modularity. The simplest way to obtain such a state is starting from an easily attainable system with a point attractor, but it has been shown that it is possible to build such a state based on its description recognized in the evolution started from the point attractor. These are 'small lakes of (nodes) activity' in 'ice' (the area of the network where nodes do not change their states) - a picture similar to the one described by Kauffman (in the network parameters space) of the 'liquid' area at the boundary of the ordered (‘frozen / solid’) area and chaotic (‘gas’). The Kauffman model is the basis of the famous hypothesis "life on the edge of chaos", however, this model strongly limits the parameters allowed for modeling life to $K = 2$ and $s = 2$ and their immediate vicinity called phase transition. Half-chaos allows a much larger range of these parameters, many estimates indicate such a need.

The experiments used a constant random structure (mainly scale-free and Erdős-Rényi random networks, but also others), random initial states of nodes and random functions, but despite the maintenance of characteristics indicating the randomness of the function, they were non-randomly correlated with states. Evolutionary variability concerned node functions. However, half-chaos was also observed in experiments, where the network grew by random addition and removing of nodes. This testifies to the more general nature of the discovered phenomenon of half-chaos.

Acceptance (as an evolutionary change) of a disturbance that gives great damage leads to ordinary chaos, which practically does not return to half-chaos. This is the elimination model - death. On the other hand, acceptance of a disturbance that gives small damage is enough to remain in half-chaos. This feature is: 'evolutionary stability of half-chaos', it is one of the most important, added to the definition of half-chaos. It creates a natural criterion of identity of the evolving object. The distinction between small and large damage is natural because they create separate peaks in the distribution of damage size separated by a large gap, in which there are practically no counts.

The discovery of half-chaos radically changes the vision of the dynamics of the studied systems. The famous Kauffman's hypothesis 'life on the edge of chaos' is strongly reinterpreted to 'life evolve in half-chaos of not fully random systems' and the analogy to phase transition is substituted by the comparison of half-chaos to 'superheated liquid'. Strong limitations contrary to the observation, on the parameters of modeling of purposeful systems are removed.

# List of abbreviations and new terms

Abbreviations used in figure descriptions are defined in those descriptions.
***N*** **–** network consists of ***N*** nodes.
***K*** – connectivity. A node in a network receives signals at the ***K*** inputs.
***k*** - number of output links of the node.
**types** of networks ***sf*** - scale-free (Barabási-Albert), ***er*** - classic "random" Erdős-Rényi, ***ss*** - single-scale, ***sh*** and ***si*** are respectively ***sf*** and ***ss*** with 30% removal.
Parameters: network **type** together with ***s,K*** (treated as a vector) are the main variables in the simulations.
**RBN –** Random (classic Erdős-Rényi) Boolean Network.
**GRN** – Gene Regulatory Network proposed by Kauffman, based on **RBN**
***t*** – is the number of time steps from a disturbance initiation.
***tmx*** - the maximum number of counting steps
***A*** – Avalanche. The size of a change in a network function at time *t* after a small disturbance is measured by the number ***A*** of the nodes, which have a different state from the pattern network – identical, but without disturbance.
***d*** - damage ***d*** $= A/N$.
***dmx*** – maximal damage, i.e., Derrida equilibrium for chaotic behavior.
***P(d)*** or ***P(A)*** - the distribution of damage size at the time *tmx*, an especially important result.
***w*** - coefficient of damage propagation. $w = \langle k \rangle (s-1)/s$. For an autonomous network with fixed $K$, $\langle k \rangle = K$ and we can use $w = K(s-1)/s$.
**small change** - in obtained here the distribution of damage size ***P(d)*** for the particular system there are two peaks and the clear gap between them: the left of **small changes** (ordered behavior) and the right of big changes (chaotic, near Derrida balance).
***q*** - **degree of order** – a fraction of damage in the range of the "**small change**", the capacity of the left peak of ***P(d)***.
**chaotic parameters** - they make random system strongly chaotic.
**Half-chaos** - state of a not fully random system, with **chaotic parameters**, but small disturbances give the ordered reaction with a similar probability to the chaotic reaction.
**evolutionary stability of half-chaos** - the "**small change**" as a criterion of the acceptance of perturbing permanent changes creating the evolution is enough to stay in **half-chaos**. It was included in the **half-chaos** definition. (Acceptance of one change that gives a chaotic reaction leads to practically irreversible entry into normal chaos.)
**in-ice-module** – the set of connected active nodes surrounded by ice – inactive nodes (with the constant state).
**local cluster** - the set of nodes with the same period of their states in the process ended of accumulation. It corresponds with **in-ice-module**.
**global cluster** - a collection of **local clusters** very similar in terms of nodes composition in the evolution of one network.
**met#** - method #, where # is a digit 1 to 8. Separate experiments with different rules described in this article.

September 29, 2020. I like more my formatting as more easy to read, then I share this version.
Figure number conversion: here=IntechOpen marked in figures above as [number in IntechOpen].
1=1, 2=7, 3=8 , 4=2 , 5=10 , 6=3 , 7=5 , 8=9 , 9=4 , 10=6 . Please, use IntechOpen numbering in a citation.
Link to Online First version: https://www.intechopen.com/online-first/life-is-not-on-the-edge-of-chaos-but-in-a-half-chaos-of-not-fully-random-systems-definition-and-simu
Until book will be ready, suggested citation form: